\documentclass{aa}  

\usepackage{graphicx}
\usepackage{txfonts}
\usepackage{hyperref}

\usepackage{xcolor}

\usepackage{siunitx}

\usepackage{sidecap}

\usepackage{stfloats}

\usepackage{ulem}

\hypersetup{
    colorlinks = True,
}
\begin{document}

   \title{Multi-channel coronal hole detection}

   \subtitle{with convolutional neural networks}

   \author{
   R. Jarolim\inst{1} \and 
   A.M. Veronig\inst{1, 2} \and 
   S. Hofmeister\inst{3} \and 
   S.G. Heinemann\inst{4} \and 
   M. Temmer\inst{1} \and 
   T. Podladchikova\inst{5} \and
   K. Dissauer\inst{1,6}}

   \institute{
   University of Graz, Institute of Physics, Universitätsplatz 5, 8010 Graz, Austria \and 
   University of Graz, Kanzelhöhe Observatory for Solar and Environmental Research, Kanzelhöhe 19, 9521 Treffen am Ossiacher See, Austria \and
   Columbia Astrophysics Laboratory, Columbia University, 550 West 120th Street, New York, NY 10027, USA \and
   Max-Planck-Institut für Sonnensystemforschung, Justus-von-Liebig-Weg 3, 37077 Göttingen, Germany \and
   Skolkovo Institute of Science and Technology, Bolshoy Boulevard 30, bld. 1, Moscow 121205, Russia \and
   NorthWest Research Associates, 3380 Mitchell Ln, Boulder, CO 80301, USA
   \\ \email{robert.jarolim@uni-graz.at}
             }

   \date{Received 23 February 2021; accepted 29 April 2021}

 
  \abstract
   {A precise detection of the coronal hole boundary is of primary interest for a better understanding of the physics of coronal holes, their role in the solar cycle evolution, and space weather forecasting.}
   {We develop a reliable, fully automatic method for the detection of coronal holes that provides consistent full-disk segmentation maps over the full solar cycle and can perform in real-time.
   }
   {We use a convolutional neural network to identify the boundaries of coronal holes from the seven extreme ultraviolet (EUV) channels of the Atmospheric Imaging Assembly (AIA) and from the line-of-sight magnetograms provided by the Helioseismic and Magnetic Imager (HMI) on board the Solar Dynamics Observatory (SDO). For our primary model (Coronal Hole RecOgnition Neural Network Over multi-Spectral-data; CHRONNOS) we use a progressively growing network approach that allows for efficient training, provides detailed segmentation maps, and takes into account relations across the full solar disk.
   }
   {We provide a thorough evaluation for performance, reliability, and consistency by comparing the model results to an independent manually curated test set.
   Our model shows good agreement to the manual labels with an intersection-over-union (IoU) of 0.63. From the total of 261 coronal holes with an area $>1.5\cdot10^{10}$~km$^2$ identified during the time-period from November 2010 to December 2016, 98.1\% were correctly detected by our model. The evaluation over almost the full solar cycle no. 24 shows that our model provides reliable coronal hole detections independent of the level of solar activity. 
   From a direct comparison over short timescales of days to weeks, we find that our model exceeds human performance in terms of consistency and reliability.
   In addition, we train our model to identify coronal holes from each channel separately and show that the neural network provides the best performance with the combined channel information, but that coronal hole segmentation maps can also be obtained from line-of-sight magnetograms alone.}
   {The proposed neural network provides a reliable data set for the study of solar-cycle dependencies and coronal-hole parameters. Given the fast and robust coronal hole segmentation, the algorithm is also highly suitable for real-time space weather applications.}


   \maketitle
%

\section{Introduction}
\label{section:introduction}

\begin{figure}[!htb]
    \centering
    \includegraphics[width=\linewidth]{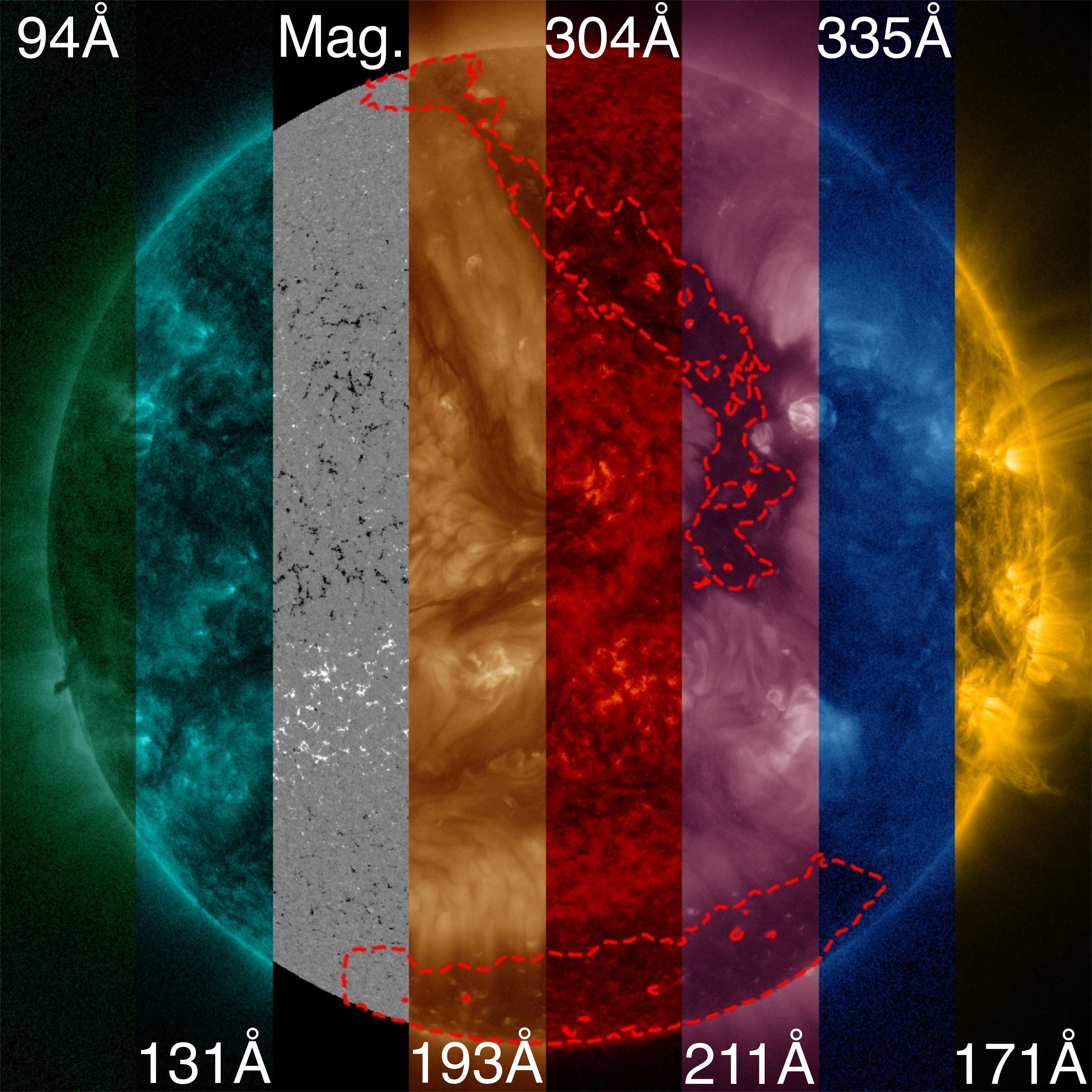}
    \caption{Composite multi-channel observation from SDO AIA and HMI on  2015-04-02. The corresponding wavelengths are given at the top or bottom of the slices. The red contour lines indicate the coronal hole boundary as estimated by CHRONNOS. The extended coronal hole in the northern hemisphere can be clearly distinguished from the quiet-Sun region in the channels 193~{\AA}, 304~{\AA,} and 211~{\AA}. Extended filament channels in 193~\AA\- show a similar dark appearance, but are correctly omitted by our method.}
    \label{fig:logo}
\end{figure}

Coronal holes appear as large-scale dark regions in extreme ultraviolet (EUV) and soft X-ray (SXR) images of the upper solar atmosphere. They are built from a large number of small-scale magnetic funnels which are rooted in the solar photosphere, expand with height throughout the chromosphere and transition region, and eventually form the "holes" in the solar corona. Coronal holes are characterized by a reduced temperature and density as compared to the ambient corona, have a dominant magnetic polarity, meaning that most of the magnetic funnels have the same magnetic polarity, and further exhibit an "open" magnetic field topology, that is, the magnetic funnels close far into interplanetary space at distances of several astronomical units \citep[e.g., review by][]{cranmer2009}. 
Solar plasma is accelerated radially away from the rotating Sun along these open magnetic field lines, forming high-speed solar wind streams which transcend the heliosphere with velocities of up to \SI{800}~{km~s$^{-1}$}. Whenever these high-speed streams hit Earth, they interact with the Earth's magnetosphere and can cause geomagnetic storms and substorms \citep[e.g., review by][]{tsurutani2006}. 


Solar coronal holes are usually identified from (1) EUV or X-ray images reflecting the density distribution of highly ionized ions in the solar corona \citep{vrsnak2007, rotter2012relation, krista2009, verbeeck2014spoca, lowder2014measurements, garton2018, illarionov2018segmentation, hamada2018automated}, (2) He~I~\SI{10830}~{\AA} images probing the solar chromosphere \citep{henney2005, tlatov20014, webb2018}, or (3) magnetic field extrapolations \citep{pomoell2018, pizzo2011, asvestari2019, jeong2020aipfss}. In EUV and soft X-ray images, coronal holes appear as distinct dark features because of their reduced density and temperature. However, the automated identification of coronal holes in these images from classical methods remains challenging, for several reasons: (a) The brightness of the solar corona varies strongly over the solar cycle \citep{brueckner1981, huang2016, heinemann2019statistical}, (b) the corona is an optically thin medium, that is, the pixel intensities give the line-of-sight integrated emission over large column depths (heights) and therefore do not relate to a well-defined atmospheric layer; and (c) other features, such as filament channels, also show reduced EUV and SXR intensities compared to the ambient corona \citep{reiss2015improvements, delouille2018}. Figure \ref{fig:logo} shows a composite of multiple EUV filtergrams and the line-of-sight magnetogram, which illustrate the wavelength-dependent appearance of coronal holes and their similarity to solar filaments.

In the chromospheric He~I~\SI{10830}{\angstrom} line, coronal holes show a slightly reduced chromospheric absorption of the underlying continuum emission, and thus appear as brighter structures in the associated He~I~\SI{10830}{\angstrom} filtergrams \citep{avrett1994formation, brajvsa1996helium}. However, the contrast is typically weak, challenging the automated detection of coronal holes in He lines. Global magnetic field extrapolations, on the other hand, identify coronal holes as the regions with an open magnetic field configuration at a given source surface height \citep[typically at $2.5$ solar radii;][]{schatten1969, levine1977, wang1990}. As the magnetic field distribution cannot be observed on the back side of the Sun and measurements of the polar magnetic fields suffer from increased uncertainties, the resulting boundaries are at best good estimates. To date, 
there exists no algorithm able to robustly and automatically identify coronal holes over long portions of a solar cycle. Nevertheless, such an algorithm would provide a far better understanding of the physics of coronal holes and their role in the solar cycle evolution, and could be used for space weather forecasting.

Here, we tackle the problem of automated coronal hole detection with state-of-the-art technology in semantic image segmentation. We train a convolutional neuronal network (CNN) to identify the boundaries of coronal holes from the seven EUV channels of the Atmospheric Imaging Assembly (AIA) and from the magnetograms taken by the Helioseismic and Magnetic Imager (HMI) on board the Solar Dynamics Observatory (SDO). The line-of-sight (LOS) magnetograms measure the field in the solar photosphere, whereas the seven EUV channels probe the solar plasma at chromospheric and coronal layers, giving an extensive view of the outer solar atmosphere. Neural networks have the ability to learn directly from multi-dimensional data and can identify coronal holes based on their shape, structural appearance, global context information, and multi-wavelength representation. This allows our neural network to efficiently and reliably distinguish between coronal holes and other dark regions (e.g., filaments). Based on the simultaneous use of six coronal EUV channels, chromospheric He II observations (304~{\AA}), and photospheric LOS magnetograms, we provide the first method that  simultaneously incorporates all the different relevant information considered by classical coronal hole detection methods.


\section{Method}
\label{section:method}

In this study, we address the problem of coronal hole detection with the use of CNNs, which are a special type of deep learning architecture well suited to high-dimensional data, such as images \citep{lecun2015deep}. A CNN consists of layer-wise convolution operations, where each layer has multiple filters (or kernels) that each consist of a set of learnable weights. For each filter, a feature map is computed by applying a convolution between the feature maps of the previous layer and the filter, where the first feature maps are the channels of the input image (e.g., red-green-blue for a regular image) and the last feature map resembles the output \citep{goodfellow2016deep}. Therefore, the set of feature maps at each layer corresponds to a local representation within the CNN. With the use of a given set of input--output pairs and the backpropagation algorithm, the weights of the neural networks can be adjusted to provide a mapping between the samples (supervised training). With a sufficiently large data set, a general mapping can be found that is also valid for novel data (generalization). At every layer, the local field of view is defined by the size of the filter, that is, the spatial extent where local features are correlated. In order to increase the field of view of the neural network, deeper architectures, down-sampling operations, and dilated convolutions are used \citep{long2015fully, yu2015multi, chen2018encoder}. Convolutional neural networks take advantage of the typically high correlation of local groups (e.g., adjacent pixels in an image) and hierarchically extract features \citep[edges $\rightarrow$ motifs $\rightarrow$ parts $\rightarrow$ objects;][]{lecun2015deep}. With this, deep CNNs extract, by layer-wise transformations, the feature representation of an image that serves as a basis to, for example, classify images \citep{armstrong2019fast}, predict sequences \citep{upendran2020solar}, transform images to different domains \citep{kim2019solar, jarolim2020image}, or provide segmentation maps \citep{long2015fully}.

From deep learning applications for semantic segmentation, the concept of multi-scale architectures has proven to be most successful \citep{ronneberger2015u, chen2018encoder, badrinarayanan2017segnet}. Here the spatial dimensions are successively reduced, while the depth of the network (number of filters) is being increased. With this, the neural network can correlate features of different scales at each resolution level and can solve more complex tasks with the increased number of filters (greater number of parameters). In order to correlate features across the full image, a sufficient network depth is required. For semantic segmentation, the output is a pixel-wise classification of the input image (segmentation map). Therefore, the feature representation (low spatial resolution) needs to be matched to the size of the input image. The segmentation map can be directly obtained from the deepest layer of the neural network \citep[e.g., ][]{long2015fully}, but with this approach the output typically lacks spatial detail. An alternative approach, which accounts for global relations and spatial resolution simultaneously, is the use of an encoder-decoder architecture \citep{ronneberger2015u, chen2018encoder, badrinarayanan2017segnet}. Here, the image is first transferred into a feature representation over multiple scales by the encoder, and then upsampled to the original resolution over the same number of scales by the decoder. At each scale, skip connections are used to combine the upsampled features, which comprise the full context information, with the spatial details from the encoder.

A common problem with deep networks is the vanishing gradient problem, where the parameter optimization of the neural network becomes inefficient with increasing network depth \citep{he2016deep, szegedy2015going}. In addition, when dealing with high resolutions, the training time increases exponentially, which often results in long convergence times for model training. A possible solution to this problem was developed by \cite{karras2017progressive} for the application of image generation. Here the authors used an approach where the size of the network is progressively increased. The advantage of this method is that the training with low-resolution samples is computationally efficient and allows  the deepest layers to be trained until full convergence is reached, before introducing higher resolutions. The shallowest layers (high spatial resolution) contain many fewer training parameters and therefore require only a fraction of the total number of optimization steps. With the progressively growing approach, efficient training can be performed even for images with high resolution, while still providing the necessary network depth to identify global relations and solve complex problems. In addition, the pre-trained layers already provide a steady gradient, which counteracts the vanishing gradient problem and divergences during training.

For our neural network we build upon a progressively growing architecture that combines all the EUV filtergrams and the LOS magnetograms into a single segmentation map (Coronal Hole RecOgnition Neural Network Over multi-Spectral-data; CHRONNOS). Here our primary aim is to obtain detailed segmentation maps, to account for the full image context, and to discover relations between the channels independently (Sect. \ref{section:CHRONNOS}). In addition, we evaluate the suitability of each single channel for coronal hole detection, which also provides us with an increased interpretability (Sect. \ref{section:SCAN}). We build upon the assumption that coronal hole features are present in each channel and employ a neural network that produces a segmentation map for each individual channel (Single Channel Analyzing Network; SCAN). A common criticism of neural networks is that they behave like a black box. This means that, although neural networks provide state-of-the-art performance for a wide variety of applications, the reasoning of the neural network is typically hidden. In this study, we mitigate this shortcoming by estimating the coronal hole boundary information for each channel separately and by evaluating the performance increase when using multi-spectral information.\footnote{visit the project page for recent updates: \\\url{https://github.com/RobertJaro/MultiChannelCHDetection} }

\subsection{Data set}
\label{section:data}

We use images recorded in the seven EUV filters of AIA \citep[][]{lemen2012aia} and line-of-sight magnetograms from HMI \citep[][]{schou2012hmi} on board the SDO \citep[][]{pesnell2012sdo} mission. For the supervised training, we use pixel-wise segmentation masks from \cite{delouille2018} as reference for training. The masks were obtained in a semi-automatic fashion, where the authors automatically extracted potential coronal holes with the SPoCA-CH module and manually reviewed the obtained segmentation maps in order to remove the remaining filaments and invalid extractions. The SPoCA-CH module is a modified version of the SPoCA algorithm \citep{verbeeck2014spoca} that gives more conservative coronal hole boundaries, less erroneously classified filaments, and less other artifacts as compared to the original SPoCA module \citep{delouille2018}. In this paper, we refer to this data set as SPoCA-CH. The data set contains a total of 2031 coronal hole segmentation maps for the time period between 2010-07-21 and 2017-01-01 and comprises observations from each day where a valid extraction could be obtained. Thanks to the review process, we expect no filaments in the data set, but we note that a significant number of coronal holes were not detected by the method or removed during the review process (about 7\%; see Sect. \ref{section:results}) and we observe method-dependent variations of the coronal hole boundary. We assume that the small deviations are sufficiently sparse and random, such that the neural network we develop can generalize to the task of coronal hole detection.

We apply a temporal separation of our data set, where we use observations from the last two months of each year for evaluation (test set) and the remaining observations for model training (training set). From this we obtain 1667 training samples. With this separation, an increased model performance due to memorization of similar observations can be excluded.

For each considered day we acquired the observation that is closest to 00:00 UT and has a valid quality flag. We obtained all seven EUV filtergrams (94~{\AA}, 131~{\AA}, 171~{\AA}, 193~{\AA}, 211~{\AA}, 304~{\AA}, 335~{\AA}) and the LOS magnetogram. For each EUV filtergram we applied a standard data reduction:
\begin{enumerate}
    \item Center the solar disk and rotate for solar north up.
    \item Spatial normalization of 1.1 solar radii to 512 pixels with a third-order affine transformation.
    \item Crop frame to 1.1 solar radii.
    \item Correct for device degradation \citep{barnes2020aiapy, boerner2014photometric}.
    \item Normalize the exposure time.
    \item Scale value range to (0, 1) based on a fixed data range, and crop values that are out of range (see Appendix \ref{appendix:scaling}).
    \item Apply $\text{asinh}$ stretch (see Appendix \ref{appendix:scaling}) and scale values to ($-1$, 1).
\end{enumerate}
The data reduction for the LOS magnetograms is performed similarly:
\begin{enumerate}
    \item Center the solar disk and rotate for solar north up.
    \item Spatial normalization of 1.1 solar radii to 512 pixels with a third-order affine transformation.
    \item Crop frame to 1.1 solar radii.
    \item Set off-limb pixels to zero.
    \item Scale values in the range ($-100$, $100$) Gauss linearly to \mbox{($-1$, 1)} and crop values outside the interval such that magnetic field strengths $>100$ Gauss are set to $100$ Gauss and values $<-100$ Gauss to $-100$ Gauss. This implies that we concentrate on weak fields, which are dominant within coronal holes and the surrounding quiet Sun regions, while we only neglect variations in the strong magnetic fields, which are mostly apparent in active regions.
\end{enumerate}
With this preprocessing, the images are consistent over the full data set and independent of device degradation or yearly variations due to the ecliptic orbit \citep[cf. ][]{galvez2019machine}.

The resized images (512$\times$512) have approximately a scale of 4.2" per pixel. As coronal holes appear as extended regions in EUV filtergrams, the 512$\times$512 pixels provide a sufficient amount of spatial detail to account even for small coronal holes.

In order to provide a test set with as few missing coronal holes as possible, we manually classify samples in the time period from 2010-11-05 to 2016-12-26 using the Collection of Analysis Tools for Coronal Holes algorithm \citep[CATCH;][]{heinemann2019statistical}\footnote{The CATCH suite is publicly available via GitHub (\url{https://github.com/sgheinemann/CATCH})}. For this data set, we again select the last two months of each year. We only include segmentation maps where a reliable extraction with low uncertainties could be obtained with CATCH, comprising about 60\% of the possible days. We only consider coronal hole contributions between longitudes of $[-400\textrm{"}, 400\textrm{"}]$ in helioprojective coordinates, where we can clearly identify coronal holes. CATCH uses a modulated threshold-based approach where the detection threshold is determined from AIA~193~{\AA} filtergrams by the intensity gradient perpendicular to the boundary. This is done by minimizing the area difference between similar thresholds individually for each coronal hole. As shown in \cite{heinemann2021statistical}, the obtained boundary reflects the temperature and density gradient of the coronal holes. We select all clearly identifiable coronal holes, check for filaments, and remove any erroneous extractions. From this we obtain 239 segmentation maps. We explicitly neglect samples from 2017 onward because of the increased difficulty in finding an appropriate boundary during periods of low solar activity with the threshold-based method. 

In summary, we consider 1667 samples for training and 239 samples for evaluation. For the comparison between CATCH and SPoCA-CH (Sect. \ref{section:results}), we consider 202 samples because of missing segmentation maps.

\subsection{Progressive growing neural network (Coronal Hole RecOgnition Neural Network Over multi-Spectral-data; CHRONNOS)}
\label{section:CHRONNOS}

\begin{figure}[!h]
    \centering
    \includegraphics[width=\linewidth]{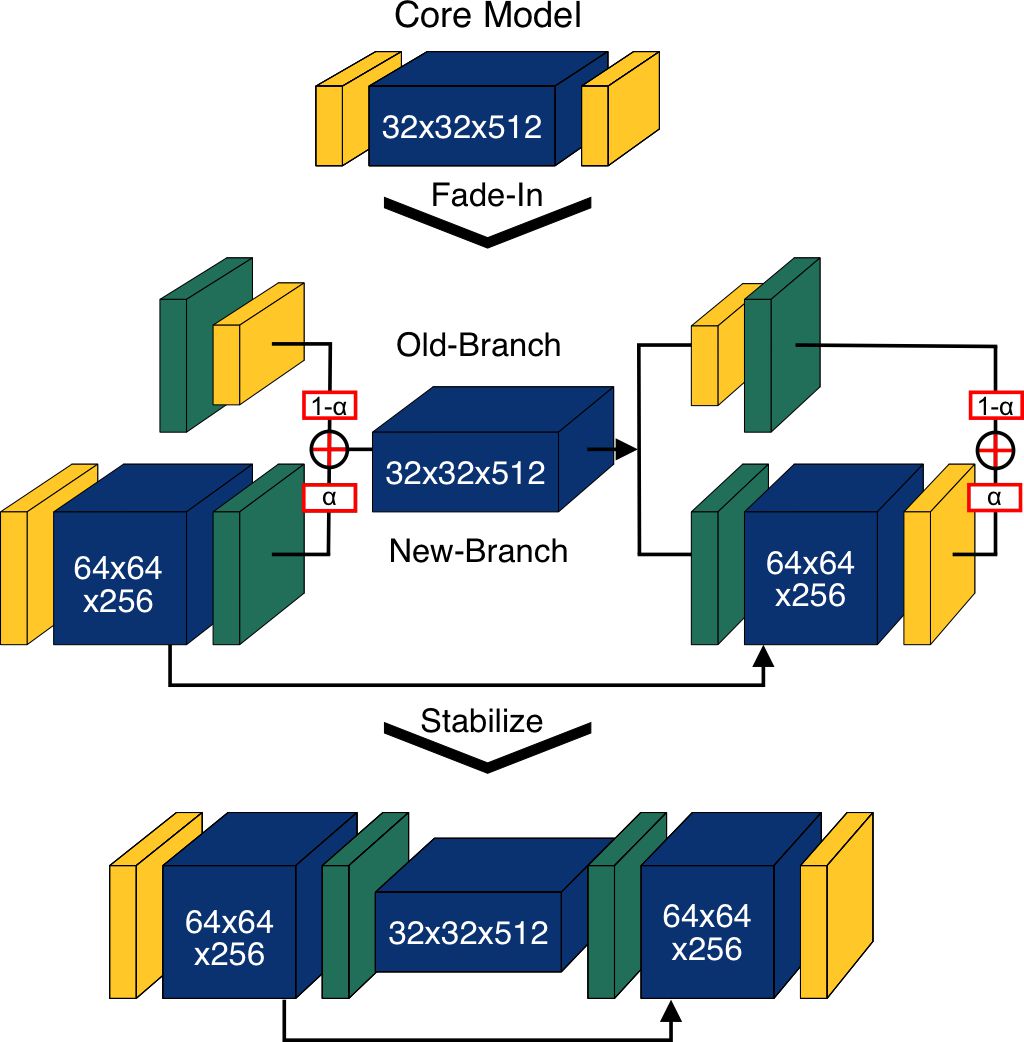}
    \caption{Training procedure of the progressively growing architecture (CHRONNOS). The next resolution level is gradually faded in by separating the input and output layers (yellow) of the original core model into a separate branch (old branch). The new branch introduces new ConvBlocks (blue) and is faded in by increasing the $\alpha$ parameter over the training cycle. The difference in resolution is adjusted by up- and down-sampling layers for the old branch and by convolutional layers for the new branch (green). Once the old branch is faded out ($\alpha = 1$), the old branch is removed and the remaining architecture is stabilized by an additional training cycle. The new branch and core model form the core model for the next resolution level.}
    \label{fig:progressive_growth}
\end{figure}

In this paper, we adapt the progressively growing image generation approach of \cite{karras2017progressive} for semantic image segmentation \citep[cf. ][]{collier2018progressively}. We employ a growing encoder-decoder architecture, where we start with an image resolution of 8$\times$8 pixels, which we progressively increase until we reach a resolution of 512$\times$512 pixels.

The majority of the trainable parameters are provided by convolutional blocks (ConvBlock; see Appendix \ref{appendix:architecture}), which we use as a central component in our neural network architecture. The training is performed iteratively in two steps, where we start with a core model that consists of a ConvBlock and an input- and output convolutional layer, which transform the input image to match the number of filters of the ConvBlock and the output of the ConvBlock to an image, respectively. In the first step, we increase to the next highest resolution by separating the old input- and output convolutional layer and introducing a new branch. The new branch consists of an input convolutional layer followed by a ConvBlock and a stride-2 convolutional layer, which corresponds to a spatial resolution decrease by a factor of two. The spatial dimension of the new input image is matched to the old input-convolutional-layer by an average pooling layer. The new input to the core model is obtained by building the weighted sum of the two branches. The output is obtained in the same way by using a transposed convolution block with stride-1/2 and an upsampling layer (see Fig. \ref{fig:progressive_growth}). A skip connection is applied between the two newly introduced ConvBlocks. The input to the core model and the segmentation mask output are obtained from the sum of both branches. Here, the new branch and old branch are weighted by $\alpha$ and $1 - \alpha$, respectively. During the training we linearly increase $\alpha$ from 0 to 1. With this gradual transition, we prevent a sudden disruption of the parameters in the core model. In the second step, we remove the old branch and train the combined architecture (equivalent to $\alpha=1$). This architecture then serves as the core model for the next iteration. Appendix \ref{appendix:architecture} contains the architecture of the fully assembled model.

The main objective of the progressively growing method is to provide high-resolution segmentation maps from the full multi-wavelength input while still being able to identify global correlations. By combining the multi-channel information at an early stage, we expect that the neural network can focus on the most relevant channels and discover relations between the channels (e.g., distinguish between filaments and coronal holes based on the underlying magnetic field information). At the deepest layer (8$\times$8 pixels), our network can correlate features across the full disk and is able to take into account the full context of the image. The intermediate blocks of the encoder-decoder architecture use skip connections that further benefit the update gradient \citep{he2016deep} and take advantage of the spatial details \citep{ronneberger2015u}. With the use of higher resolutions (512$\times$512 pixels), the network can better account for small regions (see Sect. \ref{section:results}) and provides detailed segmentation maps. The step-wise increase of image resolution provides more stable and efficient training \citep{karras2017progressive}. We note that even higher image resolutions could be achieved by further growing the architecture. In the present case, the maximum resolution of 512$\times$512 pixels was selected based on the computational limitations and the required spatial details.

\subsection{Single Channel Analyzing Network (SCAN)}
\label{section:SCAN}

For our second approach, we aim at increased interpretability. We assume that every observation channel contains information about the coronal hole boundary and train a neural network for each channel individually.

To this aim, we use the same model architecture and training procedure as introduced in Sect. \ref{section:CHRONNOS}, with the difference that we set the input dimensions to 1. We call the adapted CHRONNOS architecture Single Channel Analyzing Network (SCAN). For each channel, we train an individual neural network such that detections are solely based on the single channel information. This is different from CHRONNOS, where the detections are based on the combined representation of all channels and we cannot guarantee that each channel is equally used. Each EUV model is identical in its architecture and training. Thus, the results provide a comparison of the relative information of the individual channels for coronal hole detection. For the model training with the LOS magnetograms, we start with a resolution of 16$\times$16 pixels, grow our architecture to 128$\times$128 pixels, and use a reduced number of parameters (see Appendix \ref{appendix:architecture}) in order to counterbalance overfitting and diverging training. With the LOS magnetograms, we analyze whether coronal holes can be detected from substantially different data, where coronal hole features cannot be directly identified by humans.

\subsection{Evaluation metrics}
\label{section:metric}

\begin{figure}[!h]
    \centering
    \includegraphics[width=\linewidth]{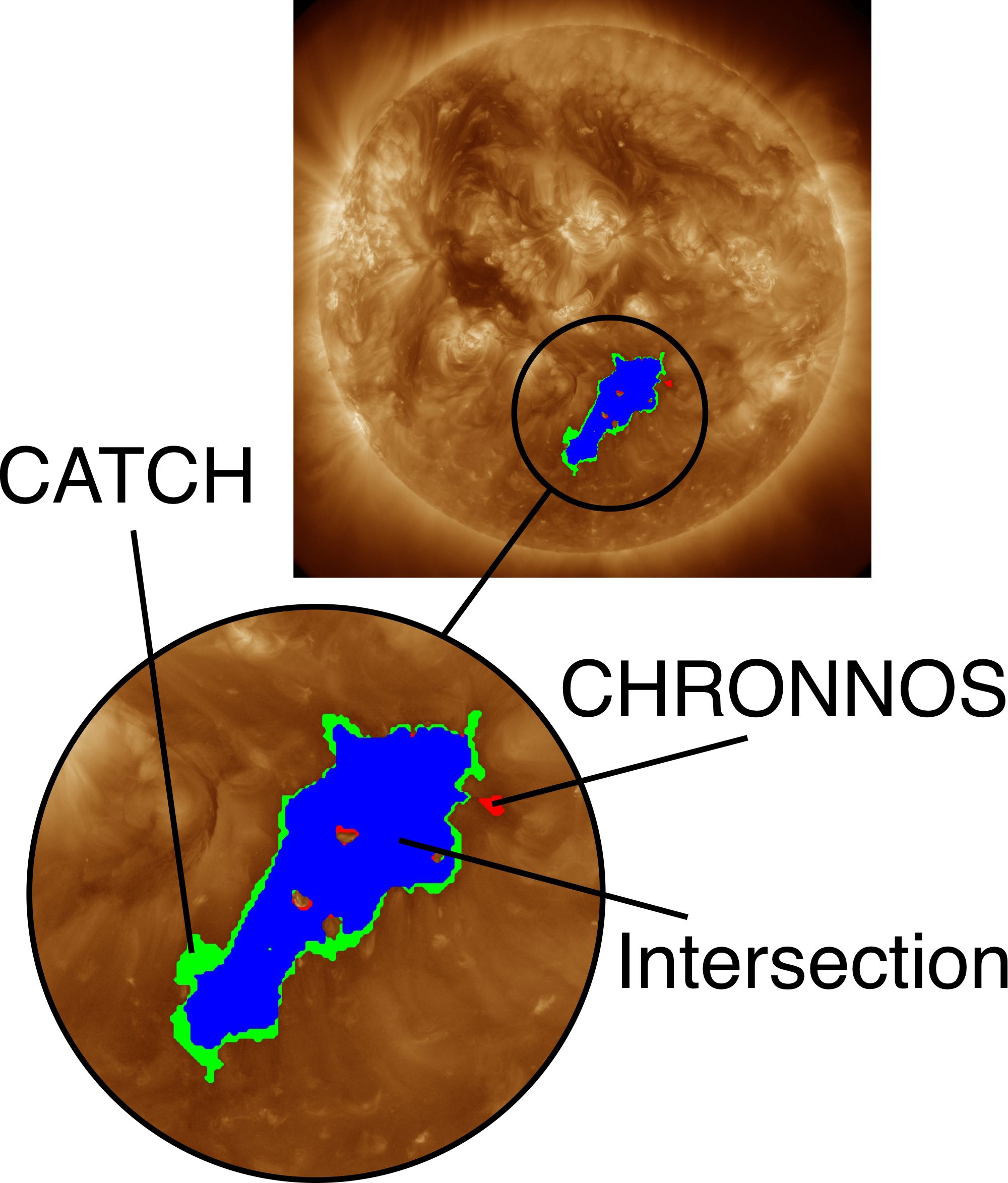}
    \caption{The IoU is computed by a pixel-wise comparison between two segmentation maps. The IoU is determined by the total number of pixels that intersect between the two segmentation maps (blue) divided by the total number of pixels  classified by either method (blue, green, and red). The example shows a slightly larger estimated coronal hole area by CATCH (green) and an additionally detected region by CHRONNOS (red).}
    \label{fig:iou}
\end{figure}

The output of the neural network corresponds to a pixel-wise probability map of the full image, with values ranging between 0 and 1. For our evaluations, we apply a threshold of 0.5 to obtain binary segmentation maps. For all our comparisons with the test set, we only consider a central section of the solar disk covering a field of $\pm400$" from the central solar meridian, in accordance with the extraction method of CATCH.

For a pixel-wise comparison of our model results with the labels from the test set, we use three metrics. We calculate the percentage of correctly classified pixels (accuracy). Because of the over-representation of noncoronal-hole pixels in the solar images, this metric will give better performance scores for conservative estimates and samples where only a small fraction of the area is occupied by coronal holes. As an alternative estimate, we use the Intersection-over-Union (IoU), which compares the fraction of overlapping coronal hole area. This metric estimates the extent to which the model detection of the coronal hole aligns with the sample from the test set, independent of the total coronal hole area in the image. An example of such a computation is given in Fig. \ref{fig:iou}, where the model prediction is shown in red, the manual CATCH labels in green, and the intersection in blue. In addition, we compute the True-Skill-Statistic \citep[TSS;][]{barnes2016comparison}:
\begin{equation}
    \label{equation:tss}
    TSS = \frac{TP}{TP + FN} - \frac{FP}{FP + TN},
\end{equation}
where $TP$, $TN$, $FP,$ and $FN$ refer to the pixel-wise estimated number of true positives, true negatives, false positives, and false negatives, respectively. The metrics are computed per data sample and averaged over the full test set.

For most applications, the individual coronal holes and their parameters (e.g., area, magnetic field) are of primary importance (e.g., space-weather, parameter analysis). \textcolor[rgb]{0.984314,0.00784314,0.027451}{\textcolor[rgb]{0,0,0}{Therefore}}, we estimate the performance of our models by determining the number of correctly identified coronal holes and by comparing the total area of the individual coronal holes. We identify coronal holes from the semantic maps by grouping coronal hole pixels, where we assign all adjacent pixels to the same group. We consider only coronal holes exceeding an area of $1.5\cdot10^{10}$~km$^2$ (corresponding to about 0.5\% of the visible solar surface). This threshold is in line with the sizes of coronal holes that show relevant correlations to high-speed solar wind streams \citep{hofmeister2018hss, tokumaru2017relation}. For each coronal hole, we determine the bounding box by the maximum horizontal and vertical extent of the grouped pixels.

The coronal hole area is determined by summing the areas (in units of km$^2$) of the labeled coronal hole pixels. To calculate the area, we first apply a geometric correction for projection effects \citep{hofmeister2017characteristics}:
\begin{equation}
    A_i = \frac{A_{i,proj}}{\cos{\alpha_i}},
\end{equation}
where $A_{i,proj}$ refers to the projected area and $\alpha$ to the heliographic angular distance from the center of the $i$th pixel. For each coronal hole we compute the full area within the bounding box in order to account for adjacent small coronal hole areas.

For a comparison in terms of individual coronal holes, we combine the bounding boxes of the prediction (CHRONNOS, SCAN, SPoCA-CH) and the reference (CATCH) by grouping boxes based on their maximum intersection, merging them by selecting the maximum extent, and removing any boxes that are enclosed within others. We evaluate the total area within each merged bounding box for both methods and classify the individual coronal hole detections. We classify coronal holes as false negative (FN) when the predicted area is below 10\% of the reference area, as false positive (FP) if the reference area is below 10\% of the predicted area, and as true positive (TP) otherwise. With this we primarily estimate the correct position of detected coronal holes, while being more tolerant in terms of coronal hole areas, which are typically identified in a more subjective manner. The detailed comparison of CH areas is given in Sect. \ref{section:ch_boundary}, including the cumulative percentage of coronal holes as function of the difference in derived areas (Fig. \ref{fig:area_difference}). 

From the evaluation of individual coronal holes, we calculate the recall,
\begin{equation}
    RCL = \frac{TP}{TP + FN},
\end{equation}
the precision,
\begin{equation}
    PRC = \frac{TP}{TP + FP},
\end{equation}
and the fraction of correctly identified coronal holes (accuracy),
\begin{equation}
    ACC = \frac{TP}{T},
\end{equation}
where $T$ refers to the total number of detected coronal holes. We note that in our evaluation the number of true negative samples is zero (TN = 0), because we cannot compare the noncoronal-hole areas in the same way.

\section{Results}
\label{section:results}

\begin{figure*}
    \centering
    \includegraphics[width=\linewidth]{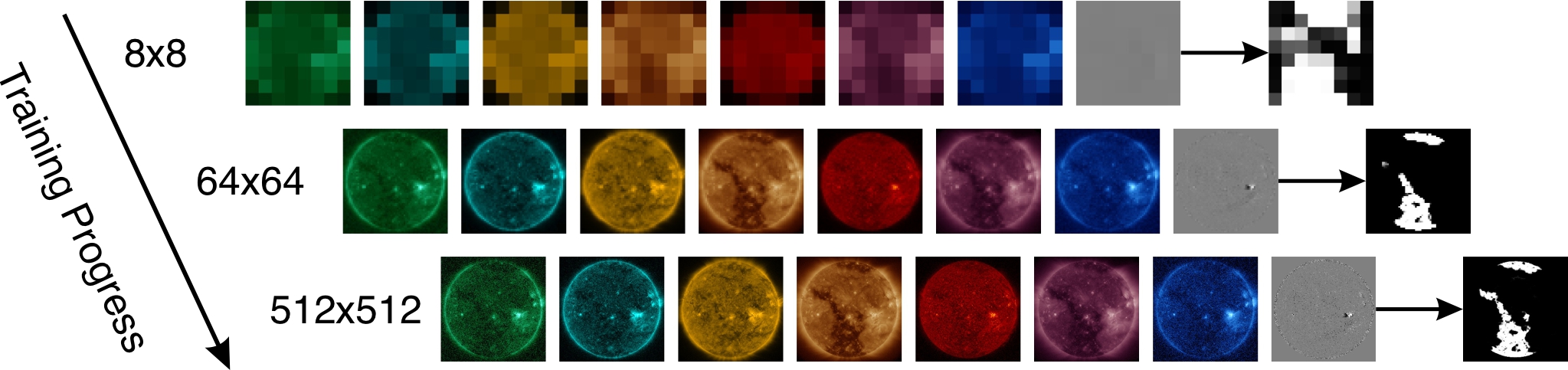}
    \caption{Example of the progressive resolution increase. The input images and labels are adjusted to the model at each resolution level. With the increase in resolution, the boundary becomes more precise and new regions can be identified that are missing at lower resolutions.}
    \label{fig:resolution_increase}
\end{figure*}

For our models, we use the Adam optimizer with a learning rate of 0.001, $\beta_1 =0$ and $\beta_2 = 0.99$ \citep[cf. ][]{karras2017progressive}. We reduce overfitting by using a weight decay of $10^{-8}$, set the dropout layer in each ConvBlock to 0.2, and augment the images by random horizontal and vertical flips. We use binary-cross-entropy as loss function and weight the noncoronal-hole class with 0.1 to account for the class imbalance. 

\begin{figure*}[!htb]
    \centering
    \includegraphics[width=\linewidth]{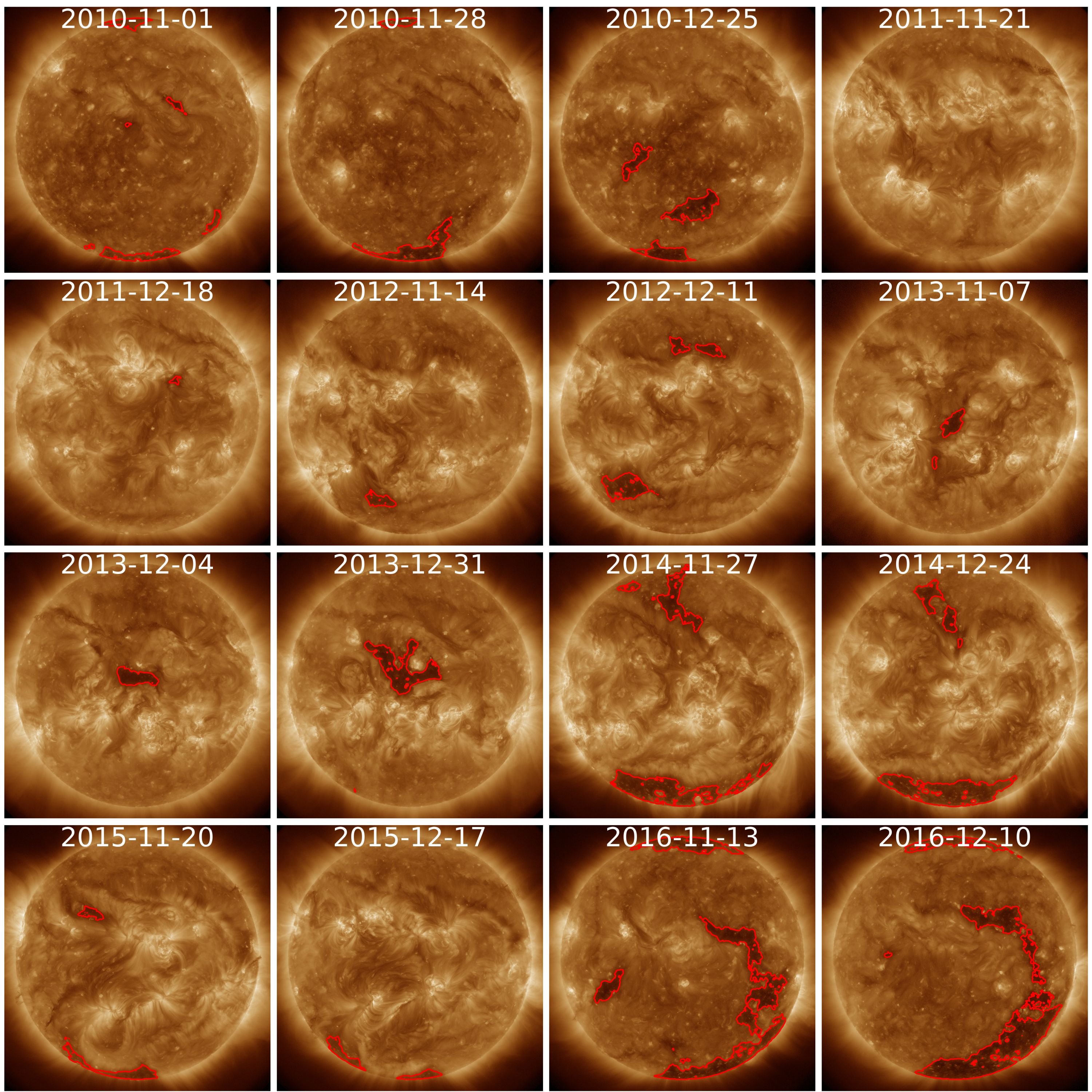}
    \caption{Samples of coronal hole detection over the test set (2010-2016; time-span of CATCH labels) of our CHRONNOS model. The red contours indicate the 0.5 threshold of the segmentation masks. The results correspond to the unfiltered output of the neural network. An animated version showing the full data set can be found online (\textbf{Movie 1}; \url{https://youtu.be/kqkjJC3eH0c}).}
    \label{fig:samples}
\end{figure*}

For the CHRONNOS model, we start with a resolution of 8$\times$8 pixels and grow the model to a resolution of 512$\times$512, as described in Sect. \ref{section:CHRONNOS}. For the resolution adjustment of the input images and the output maps we use average binning. At each resolution level, the network is trained to predict the down-sampled labels from the eight input channels (see Fig. \ref{fig:resolution_increase}). We start with a batch size of 64 and decrease it at each resolution level by a factor of two such that the batch size at the highest resolution is 1, as determined by our computational limitations. The number of training epochs is adjusted in the same way, with 10 training epochs at the highest resolution and a maximum of 100 epochs. An overview of randomly selected samples from the test set is shown in Fig.~\ref{fig:samples} and the accompanying movie, showing the results of CHRONNOS over the years 2010-2020 at a cadence of one day, can be found online (\textbf{Movie 1}; {\url{https://youtu.be/kqkjJC3eH0c}}). The red contour lines indicate the 0.5 probability threshold.

\begin{figure*}
    \centering
    \includegraphics[width=0.95\linewidth]{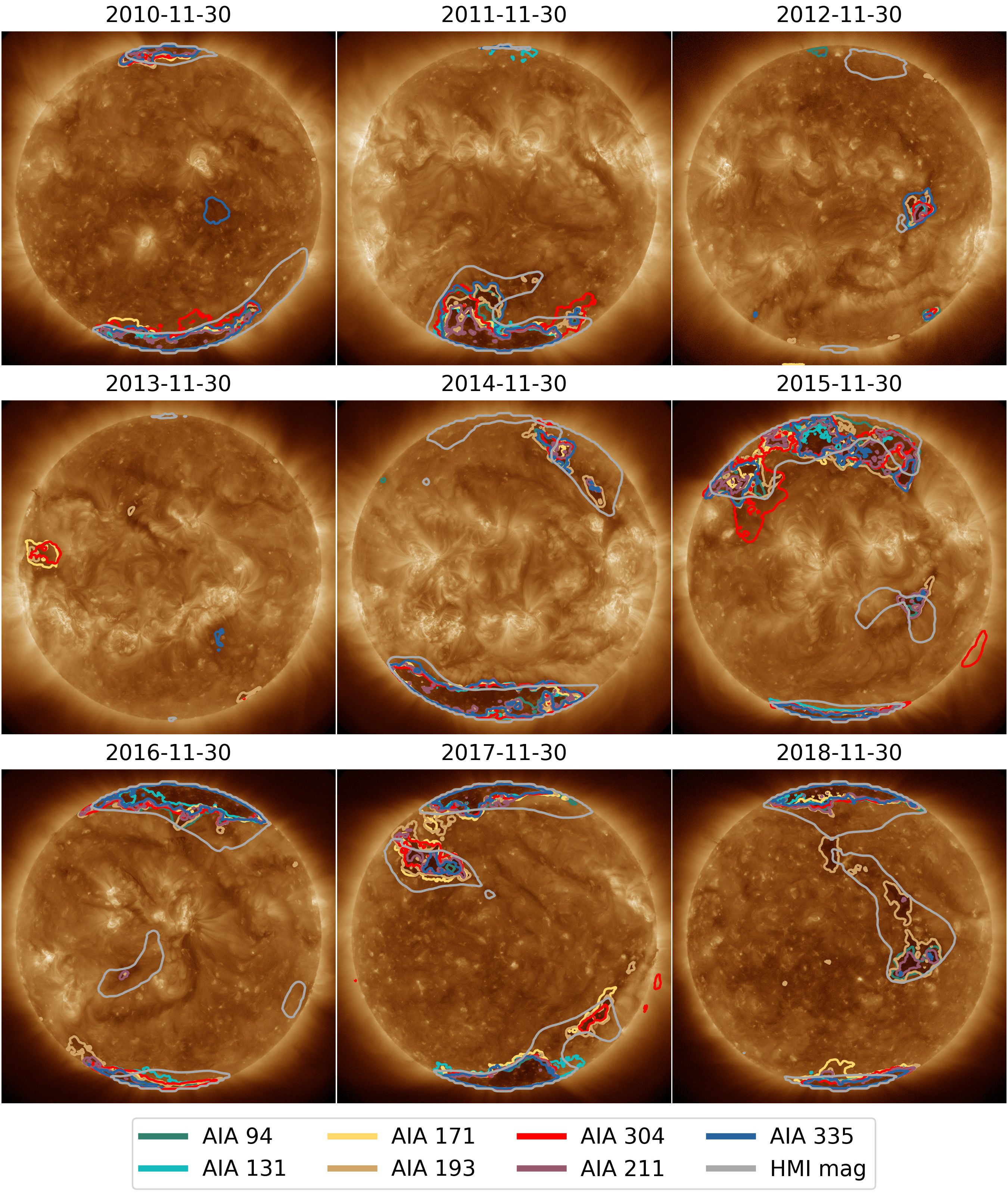}
    \caption{Comparison of the coronal hole boundary for each channel as detected by the SCAN models during 2010 -- 2018. We show one example per year, each for the day November 30 (part of the test set).}
    \label{fig:scan_samples}
\end{figure*}

\begin{figure*}[!h]
    \centering
    \includegraphics[width=\linewidth]{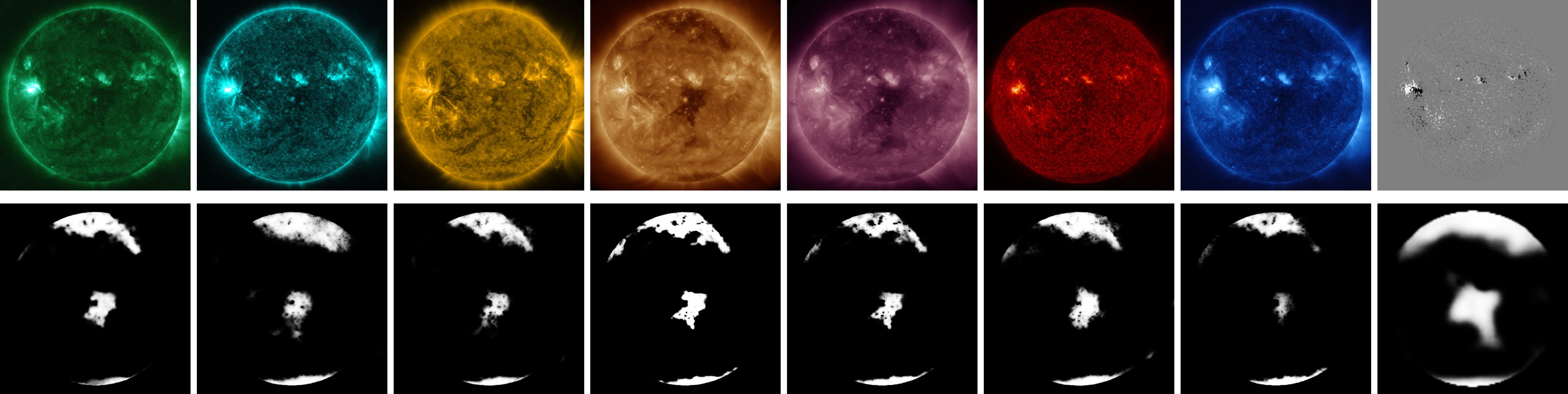}
    \caption{Sample of the segmentation maps by our SCAN models. For each channel we separately train a SCAN model to detect coronal holes. The segmentation maps of the individual channels are given below the corresponding input image. From left to right, AIA 94~{\AA}, 131~{\AA}, 171~{\AA}, 193~{\AA}, 211~{\AA}, 304~{\AA}, 335~{\AA,} and HMI magnetogram. Sharp boundaries suggest a clear identification of coronal holes (193~{\AA}, 211~{\AA}), while fuzzy boundaries suggest larger uncertainties in the detection (131~{\AA}, 171~{\AA}, 304~{\AA}, magnetogram). We note the flare event at the eastern limb, which does not affect the detections.}
    \label{fig:scan_maps}
\end{figure*}

For the SCAN models, we train an individual neural network for each channel, where we use the CHRONNOS architecture and adjust the input dimensions as described in Sect. \ref{section:SCAN}. For the EUV models, we apply an analogous training to that used in the multi-channel approach. For the LOS magnetograms we use a different setup, where we start with 16$\times$16 pixels, grow the model to 128$\times$128 pixels, and reduce the learning rate to 0.0001 in order to reduce overfitting and encourage a better convergence. The segmentation maps of the LOS magnetograms are upsampled to 512$\times$512 pixels in order to match the resolution of the original labels. In Fig.  \ref{fig:scan_samples}, samples across the full test set are shown, where the 0.5  probability threshold is given for each channel by contour lines. A sample of each channel and the corresponding segmentation maps are shown in Fig. \ref{fig:scan_maps}. From the resulting segmentations, the boundary variations and differences in the detections among the different channels can be seen. A movie of the evaluated test set can be found online (\textbf{Movie 2}; \url{https://youtu.be/lEn1DGmi2yI}).

Table \ref{table:evaluation} summarizes the performance evaluation of the individual channels of the SCAN models, the results of the CHRONNOS model, and the SPoCA-CH labels from \cite{delouille2018}, as compared to the independent manual CATCH labels. We identify coronal holes as described in Sect. \ref{section:metric} and find 261 coronal holes that exceed the $1.5\cdot10^{10}$~km$^2$ area threshold from the labels of CATCH and CHRONNOS. Here, the term ``recall'' refers to the percentage of coronal holes in the manual CATCH labels that were also detected by the method under evaluation, ``precision'' to the percentage of coronal holes that were detected by the evaluated method and are also present in the CATCH labels, and ``accuracy'' to the percentage of correct coronal hole detections.
As can be seen from Table \ref{table:evaluation}, the CHRONNOS model provides the best performance in terms of IoU (0.63) and TSS (0.81), and correctly identified coronal holes (accuracy of 98.1\%). This is in agreement with the assumption that neural networks can take multi-dimensional data into account and filter the most relevant information.
The precision score of the SPoCA-CH labels is high (99.5\%), as expected from the manual removal of filaments and invalid detections, but from the recall score it can be seen that this data set is missing about 7\% of the coronal holes that exceed the $1.5\cdot10^{10}$~km$^2$ threshold. 
The best recall is achieved by the SCAN-193 model (99.2\%), but this model  also shows a higher number of false positive detections, with a precision of 98.5\%.
For the pixel-based accuracy, the results of CHRONNOS, SCAN-193, and the semi-automatic SPoCA-CH labels are in the same range, where CHRONNOS achieves a score of 97.5\%.

\begin{table}
\caption{Comparison of the CHRONNOS, SCAN, and manually reviewed SPoCA-CH labels against the manual CATCH labels. We evaluate pixel-wise accuracy, IoU, recall (RCL), precision (PRC), and accuracy (ACC) in terms of detected coronal holes.}             
\label{table:evaluation}      
\centering                          
\begin{tabular}{l || c c c | c c c}        
\hline
 & \multicolumn{3}{c|}{Pixel} & \multicolumn{3}{c}{CH} \\
\hline
Model & ACC & IoU & TSS & RCL & PRC & ACC \\    
\hline                        
   SCAN- 94 & 96.6 & 0.50 & 0.69 & 93.0 & 95.5 & 89.1 \\
   SCAN-131 & 95.5 & 0.40 & 0.53 & 80.6 & 91.9 & 75.2 \\
   SCAN-171 & 96.2 & 0.45 & 0.61 & 88.8 & 92.7 & 83.0 \\
   SCAN-193 & 97.5 & 0.61 & 0.79 & 99.2 & 98.5 & 97.7 \\
   SCAN-211 & 97.4 & 0.59 & 0.76 & 96.8 & 98.4 & 95.3 \\
   SCAN-304 & 96.1 & 0.49 & 0.69 & 90.5 & 92.0 & 83.9 \\
   SCAN-335 & 96.5 & 0.52 & 0.74 & 94.0 & 94.7 & 89.3 \\
   SCAN-mag & 89.9 & 0.27 & 0.59 & 94.7 & 68.9 & 66.3 \\
   CHRONNOS & 97.5 & 0.63 & 0.81 & 98.8 & 99.2 & 98.1 \\
\hline                                   
   SPoCA-CH & 97.8 & 0.61 & 0.72 & 92.6 & 99.5 & 92.1 \\
\hline
\end{tabular}
\end{table}

After completing the training, all our models can operate in real-time and independent of human supervision. Our CHRONNOS model provides segmentation maps of the full solar disk within 33 ms on one graphics processing unit (GPU), and requires about 0.5 seconds on four central processing unit (CPU) cores. The results presented in this paper show the unmodified output of our neural networks (i.e., no cleaning of very small regions detected, etc.).

\section{Discussion}
\label{section:discussion}

From the comparison against the independent test set provided by CATCH \citep{heinemann2019statistical}, we find that our method outperforms the original SPoCA-CH labels, from which also the training set was separated. Although there is a significant amount of missing coronal holes in the SPoCA-CH set (about 7\%), our CHRONNOS model shows a much more reliable detection with only 1\% missing coronal holes. 
From this finding, we conclude that our neural network does not replicate the characteristics of the SPoCA algorithm that lead to the missing samples. The network shows successful generalization to the task of coronal hole detection and neglects incorrect training samples that contain partial or missing coronal hole detections. A similar behavior can be seen from image denoising with neural networks, where only noisy data were used for model training \citep{lehtinen2018noise2noise, baso2019solardenoising}. 

Among the presented models, the CHRONNOS model shows the best performance (Table \ref{table:evaluation}), by clearly providing the most reliable predictions (accuracy of 98.1\%) and showing good agreement in the detected regions with the manual CATCH labels (pixel-wise accuracy of 97.5\%; pixel-wise TSS of 0.81; IoU of 0.63). We find that the 0.8\% false positives by CHRONNOS all correspond to detections close to the $1.5\cdot10^{10}$~km$^2$ threshold at high latitudes, where the manual classification has larger uncertainties. From a manual revision of the full data set, we identified cases where the darkest parts of filament foot points are assigned with a coronal hole label ---which amounts to a few misclassified pixels---, which are irrelevant for most practical applications (e.g., extraction of individual coronal holes, assessment of the full-disk images) and have also very little effect on the evaluation metrics. With a detection rate of 98.8\%, we conclude that our CHRONNOS model reliably identifies coronal holes and distinguishes them from filaments.  
This is a strong improvement and is an important advantage of our model compared to others, for which the distinction of coronal holes  and filaments is a severe challenge. \cite{reiss2015improvements} evaluated the coronal hole detections from two algorithms using EUV images by comparing them with ground-based H$\alpha$ images from Kanzelhöhe Observatory to distinct filaments. These authors found that for both the original SPoCA algorithm of \cite{verbeeck2014spoca} and the intensity-based thresholding algorithm of \cite{rotter2012relation}, about 15\% of the identified coronal hole objects were actually filament regions.

From the preparation of our test set, we can expect a similar performance for novel data. We used daily observations, which lead to a high diversity of the data set, and applied a clear temporal split between test and training samples, which prevents the model from memorizing similar data samples. This is particularly important because of the long lifetime of coronal holes over several solar rotations \citep[e.g.,][]{heinemann2020statistical}. With the use of data samples from low and high solar activity, we optimized our neural network to provide consistent detections independent of the solar cycle and the associated highly variable EUV emission. The applied vertical flips during model training add an invariance for the global magnetic polarity of the Sun and thus our models will also provide comparable coronal hole detections for the next solar cycle.

\begin{figure}
    \centering
    \includegraphics[width=7cm]{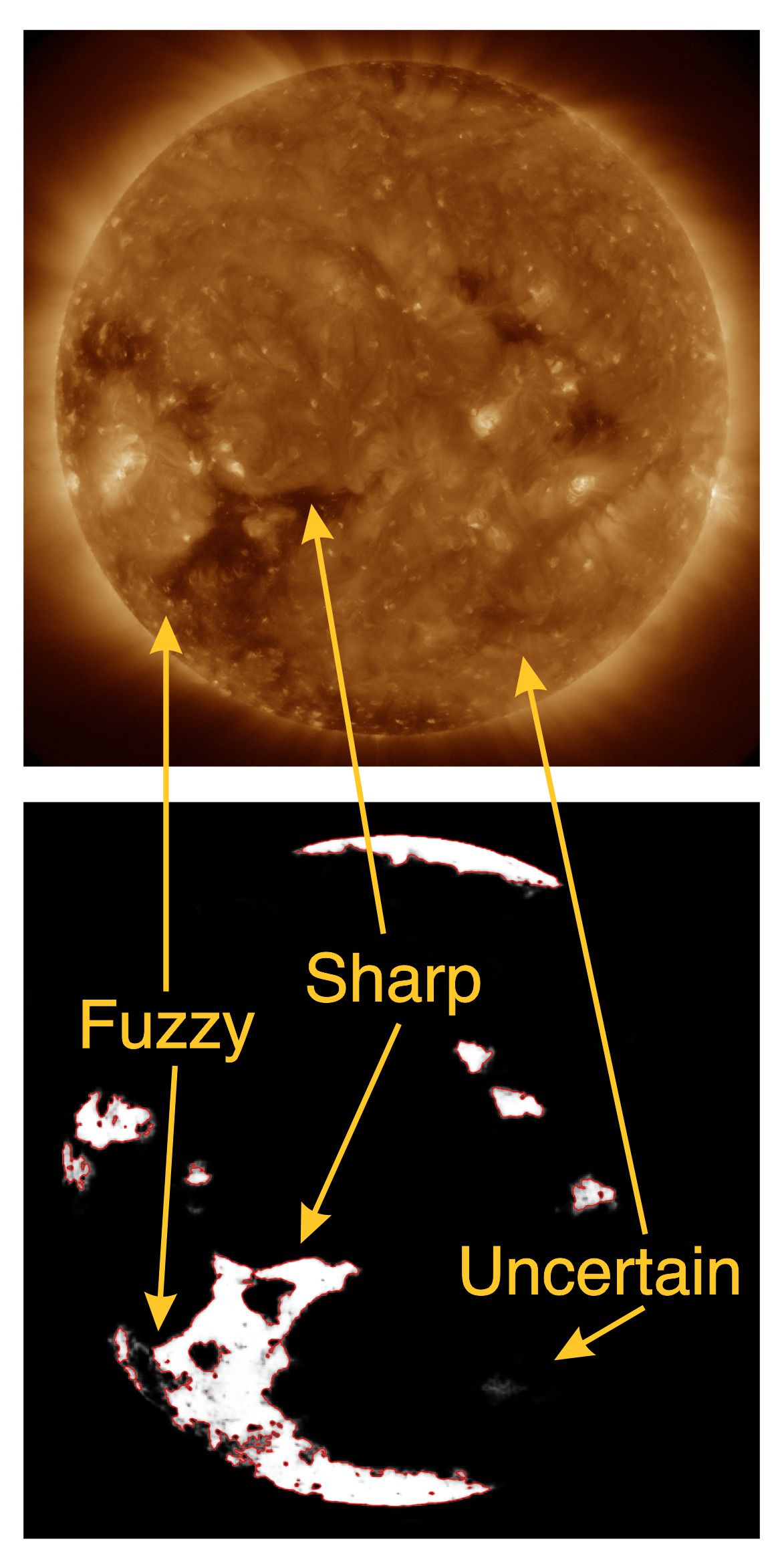}
    \caption{Example of the trained CHRONNOS model. The neural network shows high certainty at sharp boundaries, reduced probability at fuzzy boundaries, and assigns low probability values for regions that cannot be clearly classified.}
    \label{fig:probabillity}
\end{figure}

All our segmentation maps correspond to pixel-wise probabilities. Although the assignment of probability values is pixel-wise, the multi-scale architecture provides a larger view on the image and the network rather assigns probabilities to individual coronal holes and also accounts for uncertainties at the boundary. An example of a probability map of our CHRONNOS model is shown in Fig. \ref{fig:probabillity}. The neural network provides a human-like assignment of probabilities: with high probability values at sharp boundaries and for extended regions, and low probabilities at fuzzy boundaries, at coronal holes close to the limb, at polar regions and for small uncertain regions. The importance of contrast can be seen from the sample SCAN results in Fig. \ref{fig:scan_maps}, where the 94~{\AA}, 131~{\AA}, 171~{\AA}, and 304~{\AA} channels and the LOS magnetogram generally produce boundaries that are  less sharp than those produced by the channels where the coronal holes are observed in high contrast (193~{\AA}, 211~{\AA}, 335~{\AA}). For the definition of the coronal hole boundary, we set the probability threshold to 0.5. The majority of probability values are either close to 0 or 1. Only 1\% for \mbox{CHRONNOS} and a maximum of 6\% for SCAN-magnetogram lie in the range between 0.2 and 0.8, which results in small performance variations with the change of threshold.

The precise magnetic topology of coronal holes is an active topic of research, where methods such as the boundary detection from EUV images or extrapolation of photospheric magnetic fields can only provide a fraction of the information and the true coronal hole boundary still remains elusive. For space-weather aspects and the statistical study of coronal holes, the estimate of the boundary is still a key component \citep{rotter2012relation, hofmeister2019photospheric, asvestari2019, hewins2020evolution3cycles}. As the definition of the boundary strongly depends on the detection method (see \cite{reiss2021comparison} for a comparison of nine different methods for CH detection) and the wavelength range used, we argue that the consistency of the coronal hole detection is decisive for further applications. For this reason, we provide evaluations  in the following sections that estimate the consistency and reliability of our CHRONNOS model. In Sect. \ref{section:ch_boundary}, we compare the coronal hole area of the manual CATCH detections and the {CHRONNOS} detection in order to identify differences and estimate the variations. Section \ref{section:solar_cycle} asserts the long-term consistency of our method, independent of a ground-truth data set. The consistency over short timescales is analyzed in Sect. \ref{section:temporal}, where we directly compare all three methods (SPoCA-CH, CATCH, CHRONNOS). We conclude the discussion with an analysis of the importance of the wavelength used in the observations and samples of coronal hole detections from LOS magnetograms (Sect. \ref{section:channel_importance}).

\subsection{Coronal hole boundary (CHRONNOS)}
\label{section:ch_boundary}

\begin{figure}
    \centering
    \includegraphics[width=\linewidth]{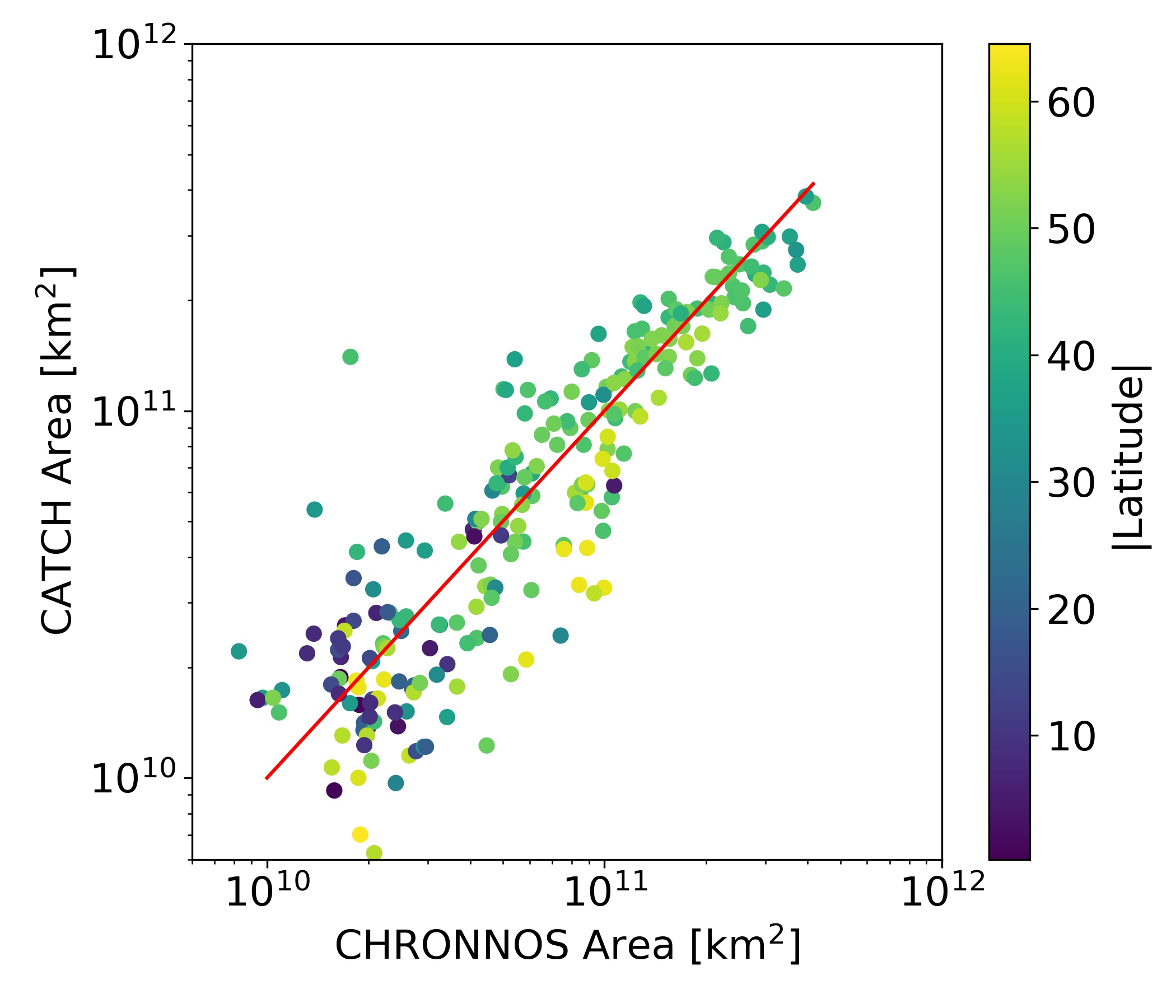}
    \caption{Comparison of the coronal hole areas as derived by CHRONNOS and the manual CATCH labels. Each data point corresponds to an individual coronal hole, where the latitude of the center of mass of the coronal hole is indicated by the color coding. The red line indicates the ideal one-to-one correspondence. The areas are plotted on a double logarithmic scale. The largest deviations occur for small and high-latitudinal coronal holes.}
    \label{fig:area_comparison}
\end{figure}

The definition of the coronal hole boundary varies between different methods and also manual extractions have a certain bias (see Sect. \ref{section:temporal} for a direct comparison). Here we compare the agreement in coronal hole area between the manual CATCH labels and the fully automatic CHRONNOS model. We compute the area of each coronal hole among the CHRONNOS and CATCH labels as described in Sect. \ref{section:metric} and plot the derived areas against each other on a logarithmic scale (Fig. \ref{fig:area_comparison}). In general, the distribution shows a good agreement between the methods, from which we conclude that our model shows no systematic over- or underestimation of coronal hole areas. The major deviations originate from small coronal holes and from coronal holes that are close to the poles. Here, projection effects play a significant role that hinders precise identification of the boundary, in particular also for manual classifications. Low-latitude coronal holes, which are the major source of high-speed stream affecting Earth and our space weather \citep[e.g., ][]{hofmeister2018hss}, show very good agreement between the methods.

\begin{figure}
    \centering
    \includegraphics[width=\linewidth]{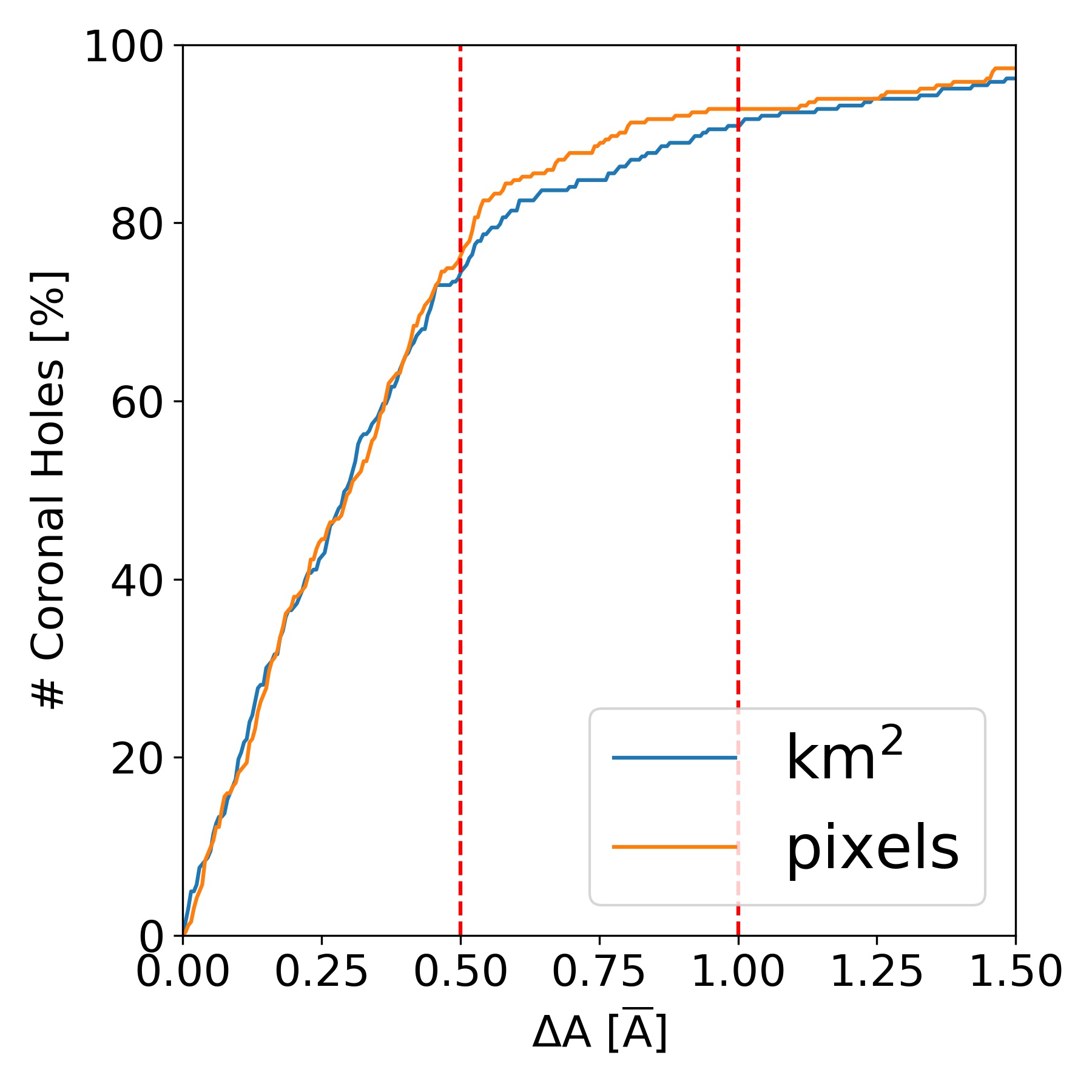}
    \caption{Evaluation of the agreement in coronal hole area. The plot shows the cumulative percentage of coronal holes as function of the difference in the derived areas $\Delta A$, given in fractions of the mean coronal hole area of the CATCH and CHRONNOS detection. The comparison is performed in terms of the deprojected area (km$^2$; blue) and pixel area (orange).}
    \label{fig:area_difference}
\end{figure}

We further evaluate the percentage of coronal holes, where the difference in area is below a given threshold. The threshold is computed as a fraction of the average coronal hole area ($\overline{\textrm{A}}$), which is determined for each coronal hole as the mean area between both methods. With this we omit the preference for a specific method and treat deviations of smaller coronal holes as equally important. In Fig.~\ref{fig:area_difference} the fraction of coronal holes is plotted as a function of the threshold. We assess our model in terms of area (km$^2$) and pixels. The slightly better performance for the pixel area originates from projection effects, which increase the difference in areas at the poles derived in km$^2$. Our evaluation shows that 75\% of the coronal holes have a difference in area of $<$50\% and that only 9\% deviate by a factor two or greater (red lines in Fig. \ref{fig:area_difference}).

\begin{figure*}[!htb]
    \centering
    \includegraphics[width=\linewidth]{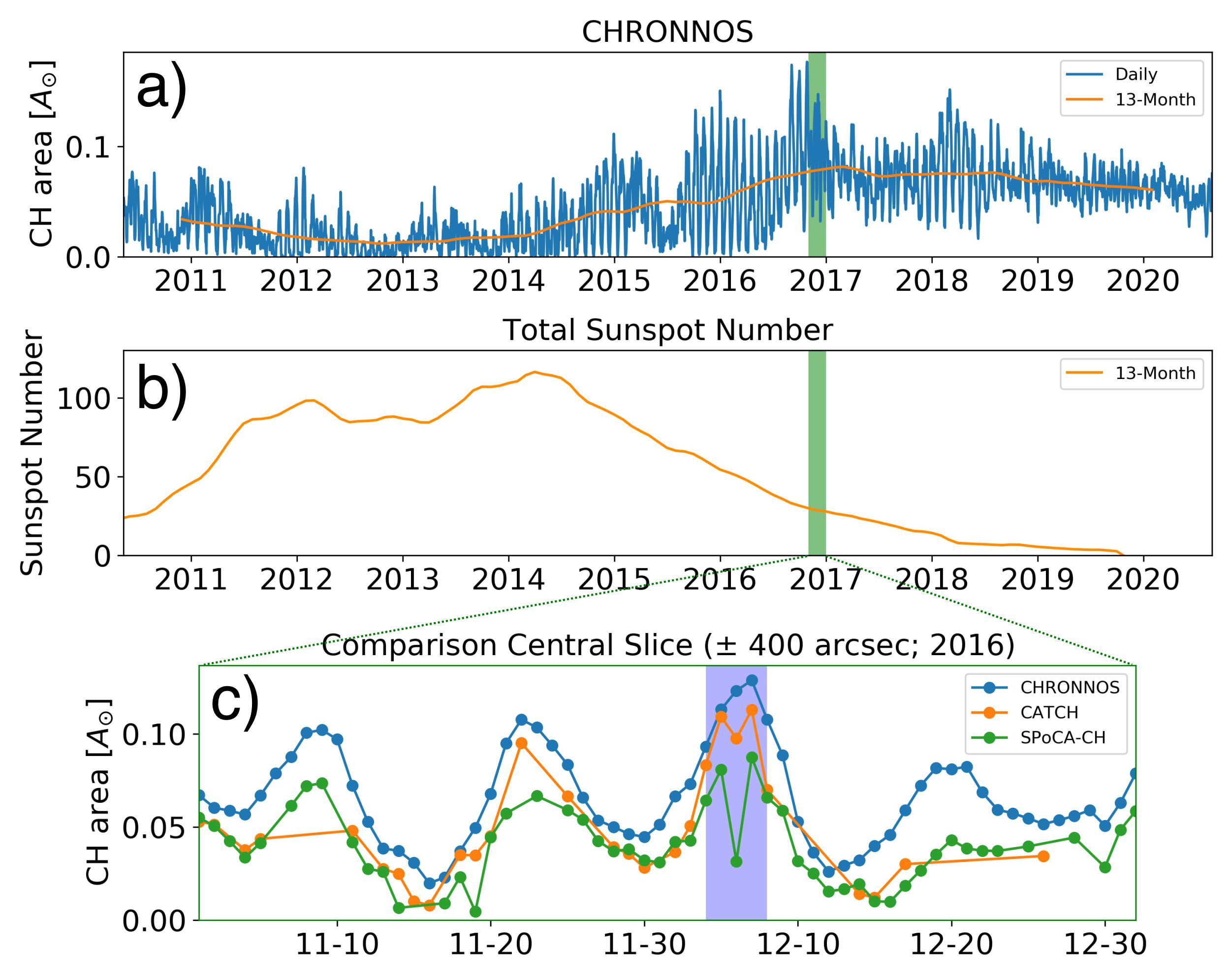}
    \caption{Analysis of the temporal consistency of the CHRONNOS model. (a) Daily variation in coronal hole area over the full disk, as given by our model, over the full data set from {2010-05-13} to {2020-08-25} (blue) and  a 13 month smoothed total coronal hole area by a mean filter (orange). (b) 13 month smoothed series of the total sunspot number, for comparison to the coronal hole area \citep[\url{http://www.sidc.be/silso/} ]{sidc}. (c) Time line from 2016-11-01 to 2016-12-31 for the CHRONNOS model (blue), the manual CATCH labels (orange), and the SPoCA-CH labels (green). The areas in panel (c) are evaluated within the central slice ($\pm400$"). The last two months of each year and all observations from 2017 onward correspond to the test set and show a smooth transition with the samples of the training set. The blue shaded area corresponds to the samples shown in Fig. \ref{fig:ch_series}.}
    \label{fig:solar_cycle}
\end{figure*}

\subsection{Solar cycle stability (CHRONNOS)}
\label{section:solar_cycle}

We estimate the stability of our model over long timescales by investigating the dependence of the total coronal hole area on solar activity. During solar maximum coronal holes can appear at all latitudes, but with shorter persistence, while during solar minimum large polar coronal holes become dominant and can be present over years \citep{cranmer2009}. From this overall relation, we expect an anti-correlation of the coronal hole area with the total sunspot number (SSN; proxy for solar activity). We use the trained model to obtain segmentation maps of each day from {2010-05-13} to {2020-08-25}, independent of previous data set assignments, and compute the total coronal hole area over the full solar disk for each observation. Figure \ref{fig:solar_cycle} shows the resulting time-series of coronal hole areas along with the international sunspot numbers for the full time range under study. We note that the last two months of each year and all results after {2017-01-01} were never considered for model training, but show a smooth transition with the remaining time-series. The short-term variations originate from the rotation signal of single coronal holes. We obtain the overall trend by applying a running mean filter of 13 months. A similar trend in total coronal hole area has also been shown in \cite{illarionov2020synopticch} using synoptic maps of AIA~193~\AA. We compare the smoothed coronal hole area of our method $x_{CH}$ to the 13 month smoothed total sunspot number $x_{SSN}$ from \cite{clette2017ssn} by computing the Pearson correlation coefficient between the two sequences, obtaining $r = - 0.88$, which shows the expected strong anti-correlation. From this we conclude that our method provides consistent coronal hole detections independent of solar-cycle variations.


\subsection{Temporal stability (CHRONNOS)}
\label{section:temporal}

\begin{figure*}[!htb]
    \centering
    \includegraphics[width=\linewidth]{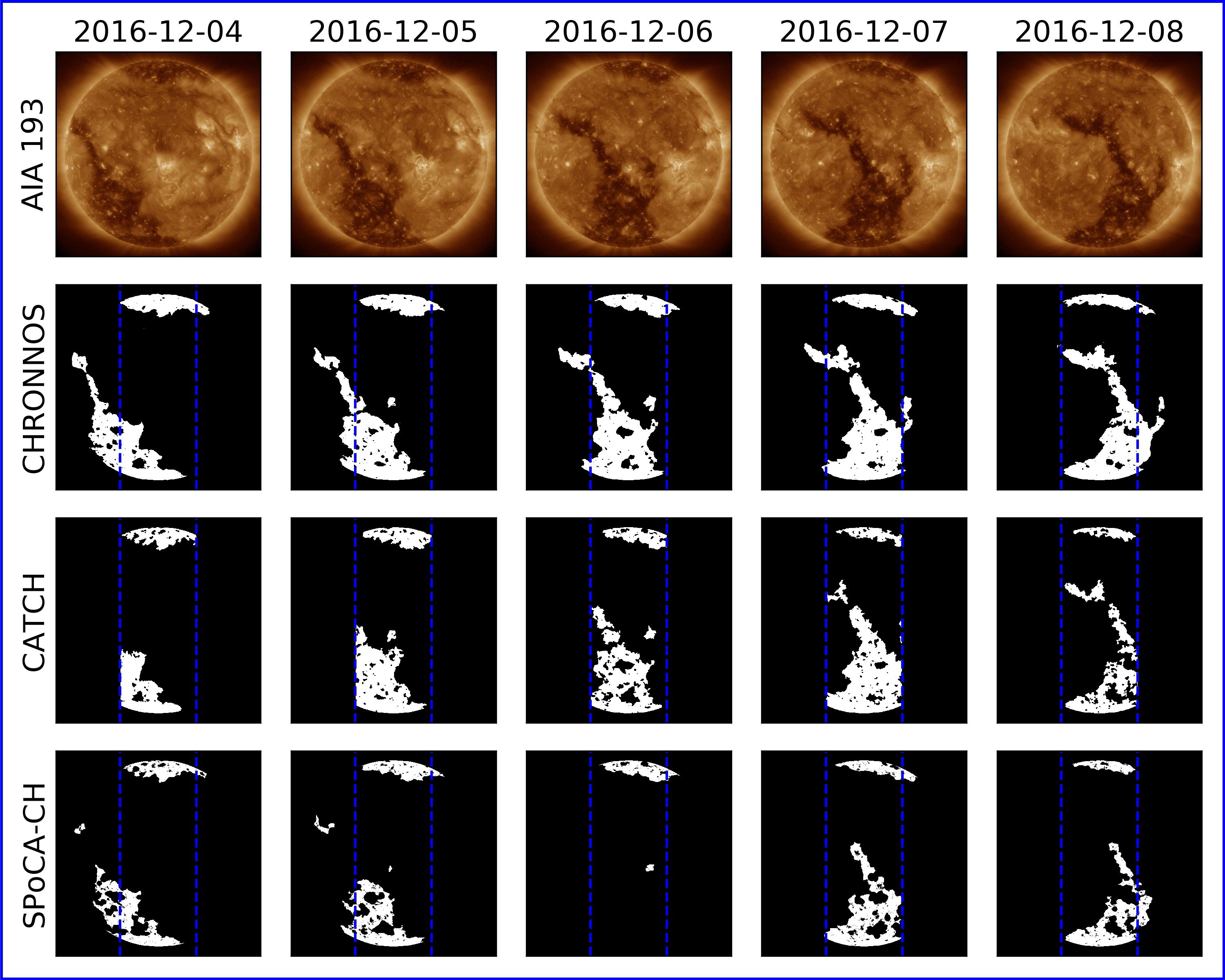}
    \caption{Comparison of the segmentation maps of the three different methods for five consecutive days. The top row shows the 193~{\AA} filtergrams of the respective day and the rows below show the corresponding segmentation maps of our CHRONNOS model, the manual CATCH labels, and the semi-automatic SPoCA-CH labels. The blue dashed lines refer to $\pm400$" of the central slice. We note that the CATCH results shown are only for the region within the slice, whereas CHRONNOS and SPoCA-CH evaluated the full solar disk.}
    \label{fig:ch_series}
\end{figure*}

In this section, we use daily observations to study the variation and coherence of the coronal hole area over short timescales, from days to weeks. From the total coronal hole area time-series shown in Fig. \ref{fig:solar_cycle}a, we select the test set samples of 2016 (green bars in Fig. \ref{fig:solar_cycle}a,b) and plot the total coronal hole area within the central slice ($\pm400$") of our neural network (CHRONNOS), the semi-automatic SPoCA-CH labels, and the manual CATCH labels (Fig. \ref{fig:solar_cycle}c). Each method operates on single observations. From the timescales of coronal hole evolution, we expect only gradual changes \citep{heinemann2020statistical}, and therefore inconsistencies in the detection would appear as sudden jumps in the time-series. 

Among the considered methods, the CHRONNOS model shows the smoothest time-series. The SPoCA-CH labels frequently show steps and sudden jumps in the time-series. Similarly, also for the manual CATCH labels we observe a lower coherence of the time-series. As can be seen from Fig. \ref{fig:solar_cycle}c, our method provides consistent detections for each observation day, without any human adjustments (see also the animated version of the full data set online \textbf{Movie 1}; {\url{https://youtu.be/kqkjJC3eH0c}}). This is an important improvement over the semi-automatic methods (SPoCA-CH and CATCH), which both show inconsistencies over daily samples and repeatedly fail to provide segmentation maps because of invalid extractions or large uncertainties.

In Fig. \ref{fig:solar_cycle}c, a discontinuity is observed on \mbox{2016-12-06} for the SPoCA-CH and CATCH detection. Figure \ref{fig:ch_series} shows the maps of five adjacent days for
comparison, which corresponds to the blue bar in Fig. \ref{fig:solar_cycle}c. The rows show the AIA 193~{\AA} image, the CHRONNOS map, the CATCH map, and the SPoCA-CH map. From a direct comparison, we can see that the CHRONNOS model is consistent over time and is in good agreement with the EUV filtergrams. The CATCH maps show slight variations in the determined boundary, which causes the discontinuity at \mbox{2016-12-06}. We note that CATCH is not designed for temporal consistency and is tuned to individual coronal holes in the center of the disk. In the SPoCA-CH maps, large coronal hole regions are even missing, which leads to the sudden drop in Fig. \ref{fig:solar_cycle}c.

From Fig. \ref{fig:ch_series} we can also compare the regions outside the central slice (blue dashed lines) between the CHRONNOS and SPoCA-CH maps. In this example, our CHRONNOS model still provides accurate and consistent predictions outside the central slice, while the SPoCA-CH method shows invalid or missing coronal hole detections towards the solar limb. This further indicates a successful generalization that outperforms the algorithm that provided the input maps for the network training.

In summary, our CHRONNOS model provides fully automatic coronal hole detections similar to the CATCH and SPoCA-CH labels, which both required manual adjustment to exclude filaments and to obtain a valid coronal hole boundary. The areas between the CHRONNOS model and the manual CATCH labels are in good agreement overall, but the comparison of daily variations suggests that the CHRONNOS detections are more consistent, robust, and reliable. As can be seen from the evaluation of the full solar cycle, the network detections appear consistent with our assumption about the long-term magnetic evolution, and also provide valid detections 
during solar minimum where other algorithms face problems. We note that all comparisons between the different methods are carried out for the central slice ($\pm400$"), but that our method also provides reasonable results close to the limb (see e.g., Fig. \ref{fig:samples} and the accompanying movie).

\begin{figure*}[!htb]
    \centering
    \includegraphics[width=\linewidth]{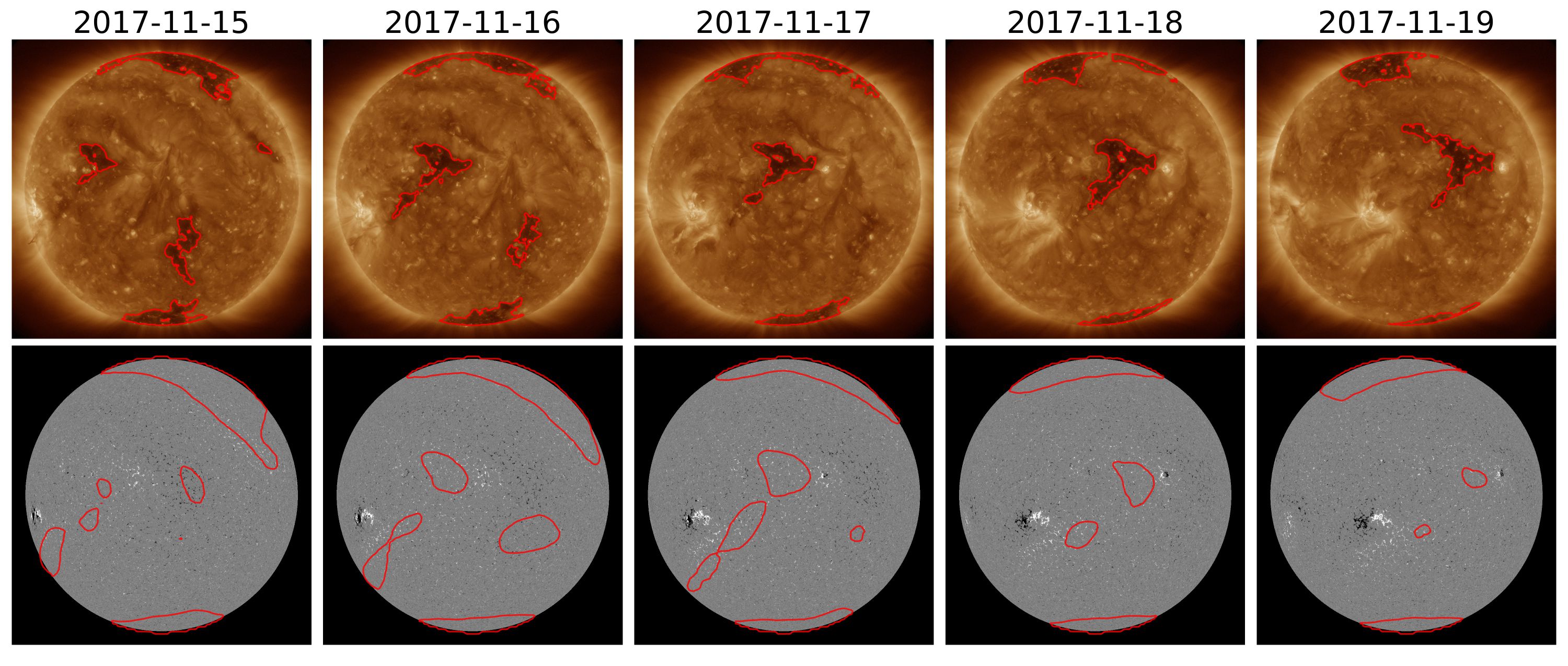}
    \caption{Comparison of CHRONNOS and SCAN-magnetogram model results of five consecutive days during solar minimum phase. The positions of the coronal holes are in good agreement, while the shapes vary significantly. An additional coronal hole close to the central active regions is detected by SCAN-magnetogram, which might be outshined in the EUV filtergram.}
    \label{fig:mag_comparison_1}
    
    \vspace*{\floatsep}

    \centering
    \includegraphics[width=\linewidth]{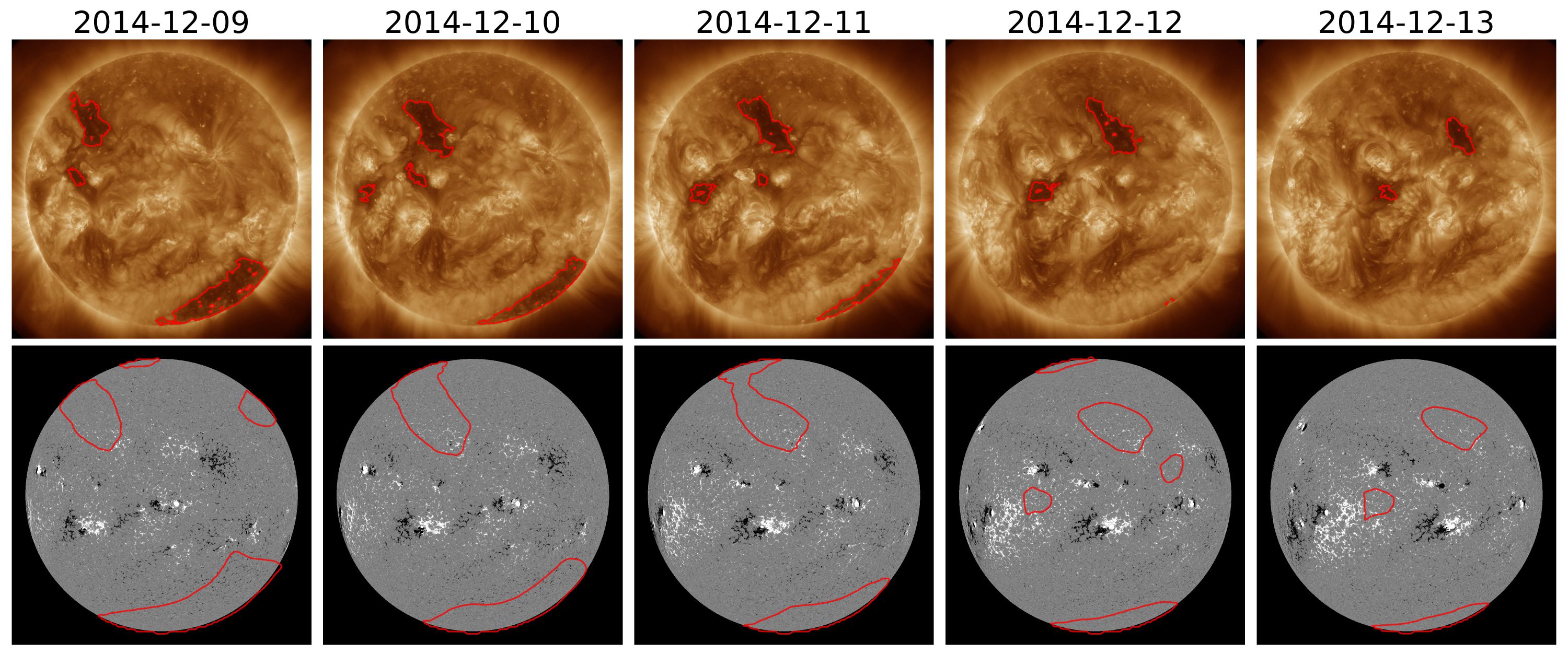}
    \caption{Comparison of CHRONNOS and SCAN-magnetogram model results of five consecutive days during solar maximum phase. The SCAN-magnetogram model shows the ability to identify even small coronal holes close to active regions.}
    \label{fig:mag_comparison_2}
\end{figure*}

\subsection{Channel importance (SCAN)}
\label{section:channel_importance}

\begin{figure*}
    \centering
    \sidecaption
    \includegraphics[width=\linewidth * 2/3]{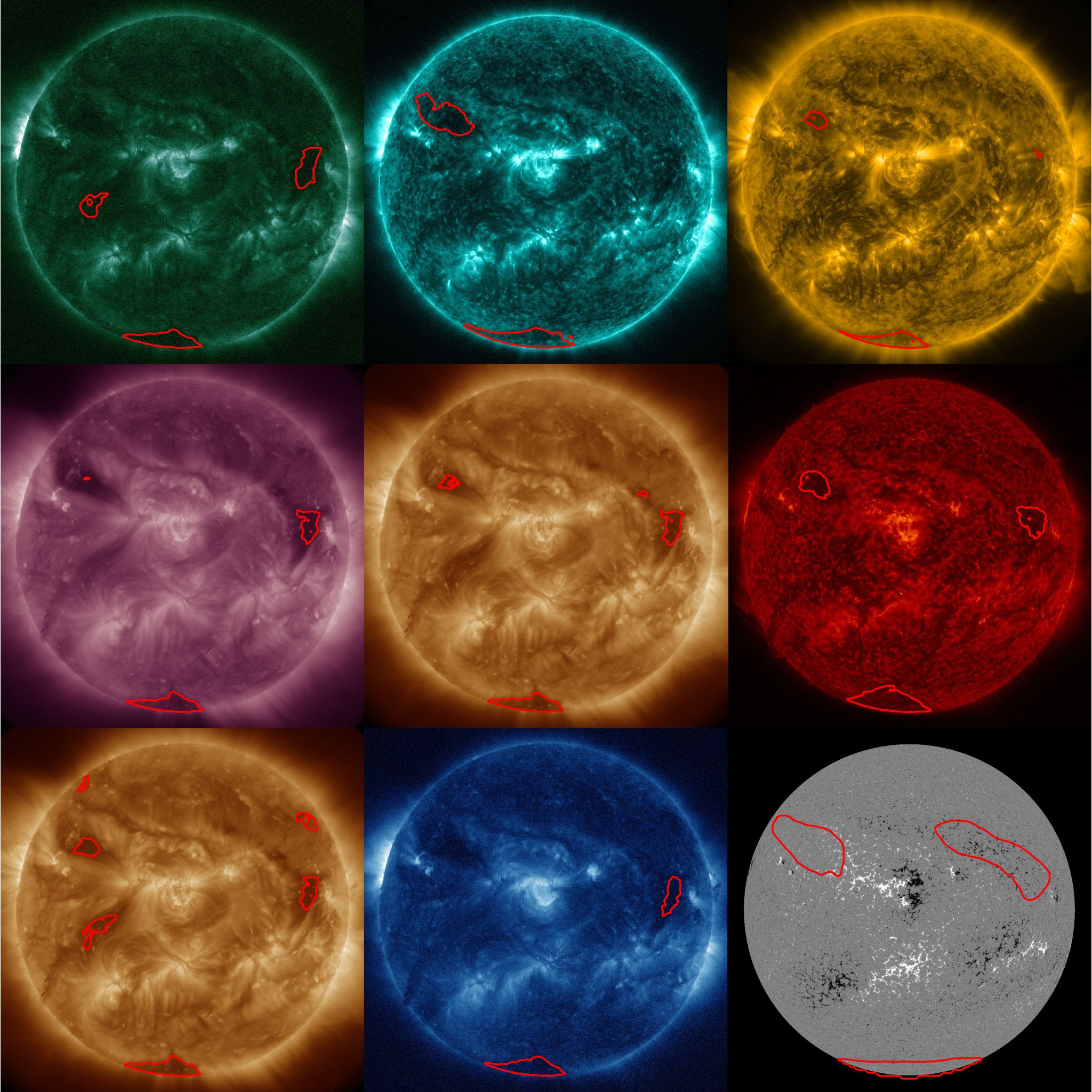}
    \caption{Sample of the SCAN models and CHRONNOS from 2015-11-19. The rows show images of (from left to right): (1) 94~{\AA}, 131~{\AA}, 171~{\AA}; (2) 211~\AA, CHRONNOS, 304~\AA; (3) 193~\AA, 335~\AA, and the LOS magnetogram. The model detections are shown as red contour lines for the corresponding channels. The combined channel result of CHRONNOS (center) is plotted as an overlay onto 193~\AA\  but is obtained from the combined set of channels. The animated version of this figure can be found online (\textbf{Movie 2}; \url{https://youtu.be/lEn1DGmi2yI}).} 
    \label{fig:scan_sample}
\end{figure*}

In order to obtain more information about the importance of each channel for coronal hole detection, we use the SCAN models that allow us to obtain segmentation maps for each channel separately. Although we can not fully investigate the decisions of the neural network, the comparison of the SCAN maps allows us to estimate the importance of the different input channels. From the similar architecture and training of the SCAN models and CHRONNOS, we can assume that differences in the results are due to the available information in the corresponding input data.

From the samples in Fig.~\ref{fig:scan_samples} and segmentation maps in Fig.~\ref{fig:scan_maps}, it can be seen that the accuracy of the boundary depends strongly on the input data. The 193~{\AA}, 211~{\AA} and 335~{\AA} filters, give the best single channel detections, with an IoU of 0.61, 0.59 and 0.52 and a TSS of 0.79, 0.76 and 0.74, respectively. This is an expected result, since the training and test labels are obtained from the 193~{\AA} channel, but is also in agreement with the different temperature sensitivities. The 193~{\AA} filter has two peaks at its response function, at 1.6~MK and 20~MK. The high temperature peak is only relevant for flares, whereas the strong signal observed by the 193~{\AA} filter at 1.6~MK samples the quiet Sun corona. Applying the AIA differential emission measure analysis, \cite{saqri2020differential} showed that the typical temperatures in coronal holes are 0.9 MK, whereas the surrounding quiet corona shows temperatures in the range of 1.5-2.0~MK. This means that the 193~{\AA} filter nicely reveals the signal in the quiet ambient corona and also provides a good contrast to the lower temperature coronal hole regions.

The 211~{\AA}, 335~{\AA,} and 94~{\AA} filters are most sensitive to plasma temperatures around 2~MK, 2.5~MK, and 6~MK, respectively. In these filters, we observe a trend of decreasing performance (IoU of 0.59, 0.52, and 0.50, respectively) with increasing temperature (c.f. Table \ref{table:evaluation}), which may be related to the decreasing data numbers (counts) in the temperature response function \citep{lemen2012aia}. This results in a lower coronal hole contrast because of the reduced emission from quiet coronal plasma, as can be also seen from the fuzzy boundaries of the coronal holes in the EUV filtergrams and detections in Figs. \ref{fig:scan_samples} and \ref{fig:scan_maps}.

The SCAN results for 171~{\AA} and 131~{\AA} show the lowest performance among the coronal emission lines, with an IoU of 0.45 and 0.40, respectively. The 171~{\AA} filter samples plasma at a temperature of 0.6~MK \citep{lemen2012aia}. This relates to the lower temperatures observed in coronal holes (1~MK) as compared to the surrounding corona, and consequently shows a lower coronal hole contrast. The 131~{\AA} filter monitors plasma at the highest temperature (10~MK), but the second temperature peak in the response function at 0.4~MK leads to similar results to those of the 171~{\AA\ filter}. By combining all the EUV channels, our method benefits from both the high contrast of the hot channels and the high count statistic of the cooler channels.

The AIA 304~{\AA} images and the HMI LOS magnetograms are associated with the upper chromosphere and photosphere, respectively. From the IoU we can see that the boundary, as estimated from the 193~{\AA} spectral line, cannot be matched from lower atmospheric layers. For AIA 304~{\AA,} this can also be
 seen from the deviations of the coronal hole boundaries in Fig. \ref{fig:scan_samples}. However, SCAN-304 provides an IoU of 0.49, a TSS   of 0.69, and a coronal hole accuracy of 83.9\%, that is, a good performance that exceeds both SCAN-131 and SCAN-171, and provides comparable results to SCAN-94 (c.f. Table \ref{table:evaluation}). This result suggests that the chromospheric line is suitable for use in coronal hole detections.

The SCAN-magnetogram model uses a different architecture and training from the SCAN models for the EUV filtergrams. We find that the SCAN-EUV training procedure is prone to underestimating coronal holes when applied to LOS magnetograms, which we associate with the deviations between the segmentation masks and the data. We also find that the 512$\times$512 pixel magnetograms do not result in more detailed boundaries. With the reduced architecture we avoid overfitting and obtain segmentation maps at an appropriate level of resolution. The evaluation of individual coronal holes shows that 94.7\% were correctly detected, but that about 30\% are false detections. There are two interpretations of this result. First, the SCAN-magnetogram results could be less reliable and therefore frequently lead to false detections. Second, the detections could be associate to open magnetic field structures in the photosphere that are outshined by active regions in the corona, which might hamper the CH detections in the coronal EUV filters. An argument that supports the second interpretation is that, for channels that contain less coronal hole information, we find lower recall scores (e.g., 80.6\% for AIA 131~{\AA}), although this could also be due to the different model architecture. We note that the false detections also include differences in size, which can be substantial at the poles where the model generally predicts larger regions (see Fig. \ref{fig:scan_samples}).

The direct detection of coronal holes from LOS magnetograms is particularly interesting for ground-based observations, where coronal imaging is not possible, but magnetograms can be obtained on a regular basis \citep{harvey1996global}. Magnetograms are used for coronal hole detection with the use of magnetic field extrapolation \citep{pomoell2018, pizzo2011, asvestari2019, jeong2020aipfss}. Here we use a different approach, by directly estimating the coronal hole boundary from photospheric magnetograms. As can be seen from Figs. \ref{fig:mag_comparison_1} and \ref{fig:mag_comparison_2}, this is a non-trivial task for a human, but surprisingly accurate detections are obtained by the neural network. Also, the network is able to correctly identify the general magnetic configuration of coronal holes as compared to the surrounding area and to find a mapping between the magnetic field information and EUV representation of the coronal holes \citep[cf.][]{kim2019solar}.

 Figures \ref{fig:mag_comparison_1} and \ref{fig:mag_comparison_2} show examples of SCAN-magnetogram alongside the CHRONNOS detections. The top row shows the multi-channel detection of CHRONNOS and the bottom row the detection made from the LOS magnetograms only. The samples show observations from five consecutive days taken during solar minimum (2017) and solar maximum (2014). The first thing to notice about the magnetograms is that they provide a much coarser detection than CHRONNOS. As we are training our neural network with segmentation maps from EUV observations, the boundary cannot be detected precisely and the network assigns larger regions. We expect that SCAN-magnetogram adjusts the coronal hole boundary for the expansion with height of the magnetic structure, which could explain the detection of different shapes, as can be seen from the coronal hole close to the disk center in Fig. \ref{fig:mag_comparison_1}, while the position of the coronal holes is in good agreement between both methods and appears consistent in time.

The observations in Fig. \ref{fig:mag_comparison_1} show that the SCAN-magnetogram model detects an additional coronal hole close to the central active region (eastern limb on 2017-11-15). While there is no evidence for the existence of a coronal hole in the EUV observation (top row), the active region could outshine the adjacent coronal hole. The appearance near the edge of magnetically complex regions is frequently observed for low-latitude coronal holes \citep{cranmer2009}. In addition, we can see from Fig.~\ref{fig:mag_comparison_2} that our network is capable of detecting coronal holes close to active regions, which supports the detection in Fig.~\ref{fig:mag_comparison_1}. This example illustrates that further analysis with multi-point observations (STEREO) and their temporal evolution is required for verification
of such detections.

From our evaluation we find that we obtain the best results by compiling the full information into a single segmentation map (CHRONNOS; Table \ref{table:evaluation}). Figure \ref{fig:scan_sample} shows a side-by-side comparison of the detections from the individual channels and the CHRONNOS detection as a contour overlay. The single-channel detections (e.g., SCAN-193) tend to also identify regions that are not clearly associated with coronal holes, but with the additional information of the other channels in CHRONNOS, such false detections can be significantly reduced. In the comparison with the CATCH test set, the 193~{\AA} detection identified 99.2\% of the coronal holes (Table \ref{table:evaluation}), but shows a substantial number of misclassifications even for larger coronal holes (98.5\% precision). The combined information approach is less prone to spurious dark regions and provides more reliable detections, with a precision of 99.2\% and a slightly higher IoU.

\section{Conclusion}
\label{section:conclusion}

In this paper, we present a reliable, fully automatic, and fast method for the detection of coronal holes using SDO/AIA and HMI full-disk images. We apply an extensive evaluation of our primary method (CHRONNOS), in comparison to an independent manually labeled data set. We provide the first method for coronal hole detection that is thoroughly assessed for reliability and temporal stability, enables the study of large coronal hole data sets over different phases of the solar cycle, and allows autonomous monitoring of the Sun. The developed neural network offers efficient training and is designed to provide detections based on a maximum amount of spectral and spatial information.

We verified 261 individual coronal hole detections that exceed an area of $1.5\cdot10^{10}$~km$^2$ (~0.5\% of the visible solar surface) in the time-period between 2010 and 2016 and show that in 98.1\% of the cases the detection by our CHRONNOS model is correct. Our neural network shows high reliability and has successfully learnt to distinguish filaments from coronal holes, with only 0.8\% coronal hole detections that are not overlapping with the test set and that all correspond to uncertain identifications of regions close to the poles.

Our primary model (CHRONNOS) achieves an IoU of 0.63, a TSS of 0.81, and correctly classifies 97.5\% of the pixels  on average. From a direct comparison of the coronal hole areas (Sect. \ref{section:ch_boundary}) we observe that errors are mainly due to differences in the definition of the boundary between the methods. The temporal evolution over short timescales shows that our neural network provides a smooth and coherent variation across daily samples, whereas the other methods show discontinuities in the time evolution of the coronal hole areas (Sect. \ref{section:temporal}). In addition, we compared the variation of total coronal hole area (i.e., summed over the solar disk) over the full solar cycle no. 24 and find a strong anti-correlation with the total sunspot number of $r=-0.88$, in agreement with the expected long-term evolution. These findings demonstrate that our method shows high temporal consistency over long and short timescales.

Our method shows successful generalization and outperforms the manually filtered SPoCA-CH labels of \cite{delouille2018}, from which we separated our training set. A comparison of the individual detection methods shows that our method even exceeds  human performance in terms of consistency, and in addition provides reasonable detections of coronal holes close to the solar limb and during solar minimum conditions (Fig. \ref{fig:samples}, \ref{fig:solar_cycle}) that could not be reliably extracted by manual labeling using CATCH.

With the separate analysis of the different EUV channels and the LOS magnetic field map, we are able to obtain a better interpretation of the importance of individual channels for coronal hole segmentation, and find that neural networks can efficiently combine multi-channel information. We show that the detections strongly depend on the temperature peak in the AIA response function, that the combined spectral information leads to the best results, and that a coarse coronal hole detection can also be directly carried out using LOS magnetograms alone.

\begin{acknowledgements}
This research has received financial support from the European Union’s Horizon 2020 research and innovation program under grant agreement No. 824135 (SOLARNET).
S.J.H. was supported, in part, by the NASA Heliophysics Living With a Star Science Program under Grant No. 80NSSC20K0183.
The computational results presented have been achieved using the Vienna Scientific Cluster (VSC) and the Skoltech HPC cluster ARKUDA.
The authors acknowledge the use of the data from the Solar Dynamics Observatory AIA and HMI science teams.
This research has made use of SunPy v1.1.4 \citep{sunpy_software2020}, an open-source and free community-developed solar data analysis Python package \citep{sunpy_community2020}.

\end{acknowledgements}

%
%

\bibliographystyle{aa}
\bibliography{references}

\begin{appendix}

\section{Data normalization}
\label{appendix:scaling}

Table \ref{table:stretch} shows the data range and the applied stretch function that is used for the image normalization of each channel. The upper limit of the data range is determined by the average maximum-value across the full data set and the lower limit is set to zero as default. The data normalization is applied after correction for exposure time and device degradation by normalizing the data based on the value range, cropping values outside the range and applying the stretch function.

The asinh stretch is computed by:
\begin{equation}
    \hat{x} = \frac{\textrm{asinh}(x/a)}{\textrm{asinh}(1/a)},
\end{equation}
where $x$ refers to the input data, $\hat{x}$ to the scaled data, and $a$ to a constant that we set to $a=0.005$.

\begin{table}
\caption{Value range and stretch function for the data normalization of the EUV filtergrams and the LOS magnetogram.}             
\label{table:stretch}      
\centering                          
\begin{tabular}{| c || r | r | r|}        
\hline
Channel & Min & Max & Stretch \\    
\hline                        
   94 \AA & 0 & 445.50 & asinh \\
   131 \AA & 0 & 981.30 & asinh \\
   171 \AA & 0 & 6,457.50 & asinh \\
   193 \AA & 0 & 7,757.31 & asinh \\
   211 \AA & 0 & 6,539.00 & asinh \\
   304 \AA & 0 & 3,756.00 & asinh \\
   335 \AA & 0 & 915.00 & asinh \\
   Magnetogram & -100.00 & 100.00 & linear \\
\hline
\end{tabular}
\end{table}

\section{Model architecture}
\label{appendix:architecture}

A ConvBlock consists of a fixed number of 2D convolutional layers, and a 2D convolutional layer with strides. For all convolutions we use a kernel size of 3, a bias term, and reflection padding. After each convolution we apply a batch normalization, followed by a leaky-ReLU activation function with 0.2 and a dropout layer with 0.2. For downsampling ConvBlocks, we apply a stride-2 convolution at the end and use the previous layer for the skip-connection. For upsampling ConvBlocks, we apply a stride-1/2 convolution (transposed convolutional layer with stride-2) as first layer and concatenate the upsampled features to the features from the skip-connection. For downsampling convolutions we double the number of filters, while for upsampling convolutions we halve the number of filters. For the output convolutional layer we omit the normalization and apply a sigmoid activation function. In Table \ref{table:CHRONNOS} an overview of the fully assembled CHRONNOS and SCAN model is given. The fully assembled CHRONNOS model contains 67.44~M trainable parameters. The architecture of the fully assembled SCAN-magnetogram model is given in Table \ref{table:scan} and contains 1.03~M trainable parameters.

\begin{table*}[!htb]
\caption{Overview of the fully assembled CHRONNOS model. For the SCAN model the input channels are set to 1. The superscript numbers of the input- and output-tensors indicate the skip-connections. Tensor shapes are given in channels-first format. Convs refers to the number of convolutional layers in the ConvBlock.}             
\label{table:CHRONNOS}      
\centering                          
\begin{tabular}{|c || c | c | c | r | r |}        
\hline
Layer & Convs & Filters & Sampling & Input Tensor Shape & Output Tensor Shape \\    
\hline                        
   Input & - & 16 & - & (8/1, 512, 512) & (16, 512, 512) \\
   ConvBlock & 1 & 16 & down & (16, 512, 512) & (16, 512, 512)$^1$; (32, 256, 256) \\
   ConvBlock & 2 & 32 & down & (32, 256, 256) & (32, 256, 256)$^2$; (64, 128, 128) \\
   ConvBlock & 2 & 64 & down & (64, 128, 128) & (64, 128, 128)$^3$; (128, 64, 64) \\
   ConvBlock & 2 & 128 & down & (128, 64, 64) & (128, 64, 64)$^4$; (256, 32, 32) \\
   ConvBlock & 3 & 256 & down & (256, 32, 32) & (256, 32, 32)$^5$; (512, 16, 16) \\
   ConvBlock & 3 & 512 & down & (512, 16, 16) & (512, 16, 16)$^6$; (1024, 8, 8) \\
   ConvBlock & 3 & 1024 & - & (1024, 8, 8) & (1024, 8, 8) \\
   ConvBlock & 3 & 512 & up & (1024, 8, 8); (512, 16, 16)$^6$ & (512, 16, 16) \\
   ConvBlock & 3 & 256 & up & (512, 16, 16); (256, 32, 32)$^5$ & (256, 32, 32) \\
   ConvBlock & 2 & 128 & up & (256, 32, 32); (128, 64, 64)$^4$ & (128, 64, 64) \\
   ConvBlock & 2 & 64 & up & (128, 64, 64); (64, 128, 128)$^3$ & (64, 128, 128) \\
   ConvBlock & 2 & 32 & up & (64, 128, 128); (32, 256, 256)$^2$ & (32, 256, 256) \\
   ConvBlock & 1 & 16 & up & (32, 256, 256); (16, 512, 512)$^1$ & (16, 512, 512) \\
   Output & - & 1 & - & (16, 512, 512) & (1, 512, 512) \\
\hline
\end{tabular}
\end{table*}

\begin{table*}[!htb]
\caption{Overview of the SCAN-magnetogram model. The superscript number for the input- and output-tensors indicate the skip-connections. Tensor shapes are given in channels-first format. Convs refers to the number of convolutional layers in the ConvBlock.}             
\label{table:scan}      
\centering                          
\begin{tabular}{|c || c | c | c | r | r |}        
\hline
Layer & Convs & Filters & Sampling & Input Tensor Shape & Output Tensor Shape \\    
\hline                        
   Input & - & 16 & - & (1, 128, 128) & (16, 128, 128) \\
   ConvBlock & 1 & 16 & down & (16, 128, 128) & (16, 128, 128)$^1$; (32, 64, 64) \\
   ConvBlock & 2 & 32 & down & (32, 64, 64) & (32, 64, 64)$^2$; (64, 32, 32) \\
   ConvBlock & 3 & 64 & down & (64, 32, 32) & (64, 32, 32)$^3$; (128, 16, 16) \\
   ConvBlock & 3 & 128 & - & (128, 16, 16) & (128, 16, 16) \\
   ConvBlock & 3 & 64 & up & (128, 16, 16); (64, 32, 32)$^3$ & (64, 32, 32) \\
   ConvBlock & 2 & 32 & up & (64, 32, 32); (32, 64, 64)$^2$ & (32, 64, 64) \\
   ConvBlock & 1 & 16 & up & (32, 64, 64); (16, 128, 128)$^1$ & (16, 128, 128) \\
   Output & - & 1 & - & (16, 128, 128) & (1, 128, 128) \\
\hline
\end{tabular}
\end{table*}

\end{appendix}
\end{document}